%% file: main.tex
\pdfoutput=1

\documentclass[11pt]{article}

\usepackage[final]{acl}

\usepackage{times}
\usepackage{latexsym}

\usepackage[T1]{fontenc}

\usepackage[utf8]{inputenc}

\usepackage{microtype}

\usepackage{inconsolata}

\usepackage{enumitem}
\usepackage{graphicx}
\usepackage{xspace}
\usepackage{booktabs}
\usepackage{multirow}
\usepackage{subcaption}
\usepackage{newunicodechar}
\usepackage{scalerel}
\usepackage{colortbl}
\usepackage{siunitx}   
\usepackage{makecell} 
\usepackage{cleveref}

\input{macros}

\begin{document}

\title{Can Language Models Replace Programmers for Coding?

\repocod \textsc{\ours}  Says ‘Not Yet’
\thanks{To appear in the Proceedings of ACL 2025. This version is for personal use only.}}

\author{{\normalsize Shanchao Liang} \\
{\normalsize Purdue University} \\
{\normalsize liang422@purdue.edu}
\And
{\normalsize Nan Jiang} \\
{\normalsize Purdue University} \\
{\normalsize jiang719@purdue.edu}
\And
{\normalsize Yiran Hu}\\
{\normalsize Purdue University}\\
{\normalsize hu954@purdue.edu}
\And
{\normalsize Lin Tan} \\
{\normalsize Purdue University} \\
{\normalsize lintan@purdue.edu}
}

\maketitle

\input{sections/Abstract}

\input{sections/Introduction}

\input{sections/Dataset}

\input{sections/Evaluation}

\input{sections/Result}

\input{sections/Related_Work}

\input{sections/Conclusion}
\input{sections/Limitation}

\input{sections/Acknowledgement}

\bibliography{main}

\input{sections/Appendix}
\end{document}

%% file: macros.tex
\definecolor{headerbg}{RGB}{220, 230, 241}
\definecolor{rowbg}{RGB}{240, 248, 255}

\definecolor{lightred}{RGB}{255, 204, 204}

\newcommand{\newsubsubsection}[1]{\smallskip \noindent \textbf{\emph{#1}}.}
\newcommand{\code}[1]{\texttt{\small #1}} %

\newcommand{\ours}{{\textsc{RepoCod}}\xspace}

\newcommand{\projectnum}{{11}\xspace}
\newcommand{\samplenum}{{980}\xspace}
\newcommand{\crossfilenum}{{498}\xspace}

\newcommand{\crossfilept}{{50.8}\xspace}
\newcommand{\infilept}{{18.1}\xspace}
\newcommand{\infuncpt}{{31.1}\xspace}

\definecolor{lightgray}{gray}{0.9}

\newcommand{\repocod}{\raisebox{-0.5em}{\includegraphics[width=2em]{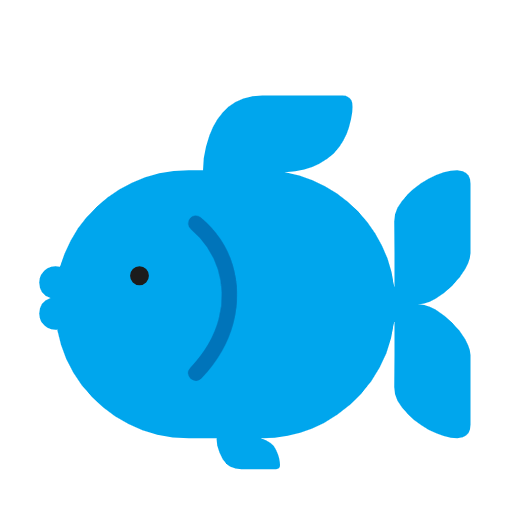}\xspace}}

\newcommand{\distance}{10pt}

%% file: sections/Abstract.tex
\begin{abstract}
Recently, a number of repository-level code generation benchmarks—such as CoderEval, DevEval, RepoEval, RepoBench, and Long-Code-Arena—have emerged to evaluate the capabilities of large language models (LLMs) beyond standalone benchmarks like HumanEval and MBPP. Thus, a natural question is, would LLMs have similar performance in real world coding tasks as their performance in these benchmarks?
Unfortunately, one cannot answer this question, 
since these benchmarks consist of short completions, synthetic examples, or focus on limited scale repositories, failing to represent real-world coding tasks.

To address these challenges, we create \ours, a Python code-generation benchmark containing complex tasks with realistic dependencies in real-world large projects and appropriate metrics for evaluating source code. 
It includes 980 whole-function generation tasks from 11 popular 
projects, \crossfilept\% of which require repository-level context. 
\ours includes 314 developer-written test cases per instance for better evaluation. 
We evaluate ten LLMs on \ours and find that none achieves more than 30\% pass@1 on \ours, indicating the necessity of building stronger LLMs that can help developers in real-world software development. In addition, we found that retrieval-augmented generation achieves better results than using target function dependencies as context.

\end{abstract}

%% file: sections/Introduction.tex
\section{Introduction}
One critical application of large language models~\cite{bian2023chatgpt, zheng2023judging} is code generation~\cite{li2023enabling, ouyang2023llm}, where  models generate executable code snippets given  natural language descriptions~\cite{starcoder,starcoder2,llama2,llama3,gpt4}. Such code generation tasks are an integral part of software development. 

Research of LLMs for code generation requires a high-quality dataset that evaluates LLMs' ability to code in real-world development scenarios. 
Existing  LLMs achieve high accuracies, i.e., over 90\% pass@1, in solving Python coding tasks in self-contained benchmarks. Would these LLMs achieve similar accuracies in real-world code-generation tasks?  To answer this question, we need benchmarks that represent real-world code generation tasks. Specifically, benchmarks need to meet the following four criteria:

Firstly, tasks should be \textbf{real-world coding tasks} 
such as %
implementing complete functions based on specifications. While manually-crafted benchmarks such as HumanEval and MBPP 
are useful at the earlier development of LLMs for code, they fail to represent realistic software development, where real project requirements drive tasks.

Secondly, \textbf{both tasks and source repositories need to be complex}, as prior work~\cite{related-dynabench-saturated-benchmark-1,related-MappingGD-saturated-benchmark-2} highlights that benchmarks quickly become saturated as models evolve and outperform them. To evaluate the true capabilities of current and future LLMs, it is crucial to construct benchmarks with challenging tasks~\cite{swebench, related-bigbench}. 
Benchmarks such as 
CoderEval~\cite{codereval}, DevEval~\cite{realted-deveval}, RepoEval~\cite{related-repocoder-repoeval},
RepoBench~\cite{repobench}, and Long-Code-Arena~\cite{longcodearena}, focus on single-line completion or short-function generation, failing to capture the complexity of multi-hundred-line functions commonly found in real-world projects. In addition, the repositories need to be complex to evaluate LLMs' ability to understand and navigate complex context. However, existing code generation benchmarks overlook this aspect, with the average number of files in their source repositories fewer than 200. Without a stronger benchmark that incorporates complex tasks and large-scale projects, we cannot reliably assess the boundaries of LLMs' code generation capabilities.

Thirdly, tasks should have \textbf{repository-level dependencies.}  Real-world code rarely exists in isolation—only 27\% of functions in 500 repositories are \textit{self-contained}, meaning they do not invoke other functions or modules within the repository and only use standard or public libraries~\cite{realted-deveval}. 
Most functions in real-world repositories depend on other components in the repository. 
Benchmarks such as CoderEval~\cite{codereval} and DevEval~\cite{realted-deveval} have the majority of tasks as self-contained or with simple dependencies, such as those dependent on the current file only.  Only 10--31\% of their tasks require dependencies beyond their current file, failing to represent coding tasks that require repository-level dependencies.

Finally, tasks should be evaluated with \textbf{appropriate evaluation metrics.} %
Similarity- or matching-based evaluation metrics, such as CodeBLEU~\cite{codebleu} and BLEU~\cite{bleu} measure textual similarities, and fail to determine whether two code snippets are functionally equivalent. 
Previous work has shown CodeBLEU and BLEU are unreliable metrics for code generation due to their high mismatch rate with human assessment~\cite{bleuisbad-outside-of-bleu}. Execution-based evaluation (e.g., unit tests), although imperfect, avoids these pitfalls by directly verifying functionality. Appendix~\ref{sec:motivating-example} uses examples to show how CodeBLEU makes mistakes. %

This paper designs and builds \emph{\ours{}}, a benchmark for assessing LLMs' ability in real-world coding tasks, i.e., generating \emph{complex} functions that require  \emph{repository-level dependencies} in real-world projects, using developer \emph{test cases} as the validation method.

This paper makes the following contributions:
\begin{itemize}[leftmargin=10pt, itemsep=0pt, topsep=-2pt]
    \item A  dataset of \samplenum challenging coding tasks from \projectnum popular Python projects, characterized by:
    \begin{itemize}[leftmargin=10pt, itemsep=0pt, topsep=0pt]
        \item Complex canonical solutions (Avg. 331.6 tokens per instance). 
         \item Extensive context from large repositories (Avg. 2,610 files per repository). 
        \item A focus on repository-level tasks—over half (\crossfilenum, \crossfilept\%) of tasks require context from other files in the repository. 
        \item Rigorous evaluation (Avg. 314 developer-written test cases per instance).
      
    \end{itemize} 
    \item A novel test collection pipeline that reduces the evaluation time to 10.4\% of the original
    \item A thorough study of ten LLMs' performance on \ours, and the key findings include:
    \begin{itemize}[leftmargin=10pt, itemsep=0pt, topsep=0pt]
        \item LLMs are ineffective on \ours{} (only 28.6\% pass@1 under oracle-retrieval).
        \item Higher recall in retrieving a target function's dependencies as context improves LLMs' performance.
        \item Providing dependencies is suboptimal for repository-level code generation tasks.
        \item For self-contained functions, additional context can still help generate better solutions.
        \item LLMs exhibit diverse strengths, as each model has uniquely solved tasks.
    \end{itemize}
\end{itemize}

%% file: sections/Dataset.tex
\begin{figure*}[t]
    \centering
    \includegraphics[width=\linewidth]{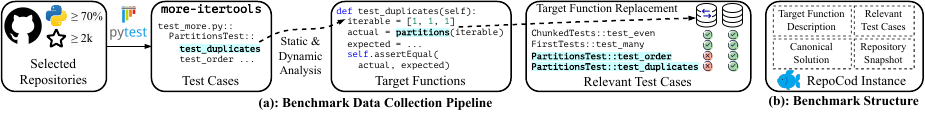}
    \caption{Data collection pipeline and instance structure of \ours{}.}
    \label{fig:overview}
\end{figure*}

\section{\ours Benchmark}
 This section introduces \ours's automated data collection process (Section~\ref{sec:data_collection}), data structure (Section~\ref{sec:benchmark_structure}) , and the statistics (Section~\ref{sec:benchmark_stat}).

\subsection{Data Collection Pipeline}
\label{sec:data_collection}
Figure~\ref{fig:overview} (a) shows the data collection pipeline. \ours utilizes GitHub as the source of our data. To filter noisy data such as functions with missing descriptions and test cases, we employ a three-stage data collection pipeline to efficiently collect target functions with good documentation and the corresponding test cases for validation. 

\newsubsubsection{Step I: Repository Selection}
The selection criteria include open-sourced repositories where Python is the primary language ($\geq$70\%) and those with no less than 2k stars, as popular repositories tend to be well-maintained~\cite{swebench}. We clone the latest version (as of October 2024) of these repositories for analysis in \textit{Step II} and \textit{Step III}.

\newsubsubsection{Step II: Target Function Selection}
For each collected repository, we perform both static and dynamic analysis to accurately identify the functions defined in the repository that are invoked by the developer-provided test cases, with detailed docstrings as function specifications. 

We first collect all the test functions in a repository using PyTest's test discovery rules\footnote{\url{https://docs.pytest.org/en/stable/announce/release-8.3.0.html}}. 
Then, we identify the functions invoked within these test functions using static and dynamic analysis techniques. For static analysis, we use tree-sitter~\cite{treesitter} to parse the test functions and collect the functions that are invoked. 
For dynamic analysis, Python's trace module is used to examine the execution of each test case. This approach also identifies the indirect function calls invoked by test functions, which are challenging to detect through static analysis alone. 

Finally, we filter the target functions, retaining only those with more than ten lines of docstrings and at least two lines of function body.

\newsubsubsection{Step III: Relevant Test Cases Collection}
To validate the correctness of LLM-generated code efficiently, we avoid running the entire test suite and instead execute only the relevant test cases—those that specifically invoke the target functions. This targeted approach significantly reduces verification time in large repositories with extensive test suites.

While the previous step provides a mapping between target functions and relevant test cases, certain mappings may still be missed. For instance, test functions might use \code{subprocess.run(command)} to invoke repository functions indirectly, which is not collected by static or simple dynamic analysis. Therefore, we employ a two-step collection procedure to capture all relevant test cases. First, we execute all test cases in the repository to establish a reference result. Then, for each target function, we replace it with an assertion failure, rerun all test cases, and compare the results to the reference. If a test result changes from \textit{pass} to \textit{fail}, it indicates that the test case is relevant to the target function.

\subsection{Benchmark Structure}
\label{sec:benchmark_structure}
Figure~\ref{fig:overview} (b) illustrates the components of \ours{}'s instances: the \emph{target function description}, \emph{repository snapshot}, \emph{relevant test cases}, and \emph{canonical solution}. \ours{} uses the developer-provided docstring as the \emph{target function description}. The \emph{repository snapshot}, is the source repository with the target function body removed. 

To use \ours{} as a code generation benchmark, the LLM is provided with the target function description and the repository snapshot, then it is expected to generate a functionally equivalent code snippet as the canonical solution. The LLM-generated solution is considered correct if it passes all \emph{relevant tests}. Finally, the \emph{canonical solutions} are the developer-written function bodies.

\subsection{Benchmark Statistics}
\label{sec:benchmark_stat}

\input{tables/dataset_stat}

\label{sec: comparison_with_existing_benchmark}
Table~\ref{table:data_stat} compares \ours{} with existing code generation benchmarks that use repository-level context, including CrossCodeEval~\cite{crosscodeeval}, RepoBench~\cite{repobench}, Long-Code-Arena (LCA)~\cite{longcodearena}, CoderEval~\cite{codereval}, DevEval~\cite{realted-deveval}, R2E-Eval1~\cite{related-r2e}, %
and RepoEval~\cite{related-repocoder-repoeval}. %

\footnotetext[1]{CoderEval only releases the task instances along with ground truths but omits the test cases.}
\footnotetext[2]{For data statistics not available in DevEval's paper, we recompute the data from their released dataset available in their GitHub repository.}
\footnotetext[3]{R2E-Eval1 has not released its dataset, and the data is derived from the published paper.}

CrossCodeEval, RepoBench, and LCA contain only single-line code generation tasks, resulting in shorter canonical solutions compared to \ours{}. In addition, they rely on similarity-based metrics, which are insufficient to evaluate code quality. Therefore, we mainly compare \ours{} against datasets that require full function generation and include test cases for evaluation, such as CoderEval, RepoEval and etc. Among all function generation datasets, \ours outperforms in nearly every aspect, with the exception of having fewer instances compared to DevEval. However, \ours leverages an automated annotation process for annotating the task descriptions for each instance, whereas DevEval relies on human effort, making \ours more scalable for expansion.

\input{tables/context_complexity}

\newsubsubsection{Repository Scale}
The instances in \ours are collected from repositories with an average of 2,610 files and 290,110 lines of code per repository. This significantly exceeds the scale of source repositories from other benchmarks, underscoring the challenge of efficiently utilizing repository-level context for the evaluated models.

\newsubsubsection{Test Scale}
\ours utilizes an average of 313.5 tests per instance, significantly surpassing other benchmarks, highlighting \ours' robustness in evaluating models. RepoEval\textsuperscript{*}'s \#Tests is computed as the average number of test cases across all available repositories, regardless of the tests' relevance to the task. If measured the same way, \ours' average \#Test would be 17,974.

\newsubsubsection{Length Complexity}
\ours presents the most challenging tasks among all datasets. With the largest average token count for canonical solutions at 331.6 (Table~\ref{table:data_stat}), LLMs must generate solutions that may require more than twice as much code as those in other benchmarks to complete the task. 

\newsubsubsection{Cyclomatic Complexity}
\ours{} also includes a cyclomatic complexity~\cite{cyclomatic} score of 9.0, indicating that the canonical solutions in \ours{} have a higher structural complexity regarding control flow.

\newsubsubsection{Context Complexity}
Table~\ref{tab:context_complexity} shows \ours, DevEval and CoderEval's distribution of tasks by three types of context complexity: \textit{Self-Contained}, \textit{File-Level}, and \textit{Repository-Level}. Self-Contained functions use only standard or public libraries; File-Level functions reference functions or classes from the same file but not others; Repository-Level functions may invoke functions or classes from the same file or other files in the repository. Among these benchmarks, \ours has the highest ratio of repository-level dependencies (\crossfilenum).

These statistics show that the tasks in \ours{} are the most challenging, and contain more repository-level context than other benchmarks, making it particularly suitable for evaluating current and future models on repository-level tasks.

\newsubsubsection{Optimized Test Execution}
With our test-collection pipeline, we reduce the number of required test cases per instance from an average of 17,974 (all available test cases from the source repository) to just 313, cutting execution time from 216.9 hours to 22.6 hours, demonstrating a substantial improvement in efficiency.

%% file: tables/dataset_stat.tex
\begin{table}[t]
\centering
\scriptsize
\setlength{\tabcolsep}{2pt}
\begin{tabular}{lrrrrr|rr}
\toprule
\multirow{3}{*}{\textbf{{Benchmarks}}}& \multicolumn{4}{c}{\textbf{Instance Statistics}} & &\multicolumn{2}{c}{\textbf{Repository Scale}} \\
\cmidrule{2-5} \cmidrule{7-8}
   & \textbf{\#Instances} &\textbf{ \#Tokens} & \textbf{{Cyclo.}} & \textbf{\#Tests} && \textbf{\#Lines} & \textbf{\#Files} \\
\midrule
CrossCodeEval & 2,665 & 13.2 & 1.0 & 0 && - & -\\
RepoBench & 23,561 & 13.7 & 1.0 & 0 && - & - \\
LCA & 934 & 12.0 & 1.0 & 0 && - & - \\
\midrule
CoderEval\footnotemark[1] & 230 & 108.2 & 4.7 & - && 48,821& 152 \\
DevEval\footnotemark[2] & 1,825 & 86.3 & 3.5 & 2.1 &&  36,640 & 164 \\
R2E-Eval1\footnotemark[3] & 246 & 127.2 & - & 11.5 && - & - \\
RepoEval\textsuperscript{*} & 373 & 84.1 & 2.7 & 2742.5 && 7,387 & 119 \\
\rowcolor{gray!20}\ours  & \samplenum & 331.6  & 9.0 & 313.5 && 290,110 & 2,610 \\
\bottomrule
\end{tabular}
\begin{flushright}
\scriptsize{RepoEval\textsuperscript{*}'s \#Tests is averaged over a subset of executable repositories, as some repositories cannot be run to collect test case counts.}

\end{flushright}

\caption{Benchmarks Comparison. `-' indicates the data is not publicly available. \textbf{\#Instances}: number of instances in each benchmark; \textbf{\#Tokens}: average number of tokens of the canonical solution; \textbf{Cyclo.} average cyclomatic complexity of the canonical solution; \textbf{\#Tests}: average number of test cases per instance; \textbf{\#Lines}: average number of lines per repositories;
}
\label{table:data_stat}
\vspace{2pt}
\end{table}

%% file: tables/context_complexity.tex
\begin{table}[t]
    \centering
    \scriptsize
    \setlength{\tabcolsep}{8.5 pt}
    \begin{tabular}{l rrr}
        \toprule
        \textbf{Dataset} & \textbf{Repository-Level} & \textbf{File-Level} & \textbf{Self-Contained} \\
        \midrule
        CoderEval & 10.0 & 53.5 & 36.5 \\
        DevEval & 31.3 & 41.2 & 27.5\\
        \textbf{\ours{}} & \textbf{\crossfilept} & \infilept & \infuncpt \\
        \bottomrule
    \end{tabular}
    \caption{Data context complexity distribution (\%) of code generation benchmarks.}
    \label{tab:context_complexity}
\end{table}

%% file: sections/Evaluation.tex
\section{Experiment Setup}
Due to the large repository sizes, most LLMs face context window limitations that prevent them from processing all the context in \ours{}. To address this, we employ Retrieval-Augmented Generation (RAG). We evaluate three popular retrieval settings: \textit{{RAG\textsubscript{BM25}}}, \textit{RAG\textsubscript{Dense}}, and \textit{current file}.
In addition to a  \textit{Baseline} (no context) setting, we use two unrealistic ``oracle'' settings, \textit{Callees}, and \textit{RAG\textsubscript{Dense-oracle}}, to study the potential best scenarios.

\subsection{Retrieval Settings}

We first define our retrieval corpus, which is used across all RAG methods. The retrieval corpus contains complete function definitions, each including the function signature and full body, sourced from all functions in the repositories.%
For \textit{RAG\textsubscript{BM25}} and \textit{RAG\textsubscript{Dense}}, since the target function body is unavailable during inference, we query using its signature and docstring to retrieve similar functions.

\newsubsubsection{RAG\textsubscript{BM25}}
We use BM25~\cite{bm25} to extract relevant functions from the corpus as context for generation. The signatures, docstrings, file paths, and bodies of the retrieved functions are provided as context.

\newsubsubsection{RAG\textsubscript{Dense}}
We encode all functions in the corpus using the \code{text-embedding-3-small} model. The target function's signature and docstring are encoded into an embedding to be compared against the function embeddings in the corpus. The top-ranked functions based on cosine similarity are provided as context, in the same format as sparse retrieval.

\newsubsubsection{Current File}
In this setting, the context is limited to the file containing the target function, with the entire file provided as context, excluding the target function's body.

\subsection{Additional Settings}
\newsubsubsection{Baseline}
In this setting, the models are provided only with the function signature and the docstring to generate the target function. This serves as a minimal context setting, evaluating the model's ability to generate functions without context information.

\newsubsubsection{Callees}  
Functions invoked by the canonical solution are considered as oracle context in several benchmarks~\cite{codereval, realted-deveval}. We extract these invoked functions and include them along with the last 1,024 tokens from the target function’s prefix. This setup evaluates whether explicitly providing referenced functions, with the target function’s most recent context, improves generation quality.

\newsubsubsection{RAG\textsubscript{Dense-oracle}}
We include the canonical solutions along with the task descriptions as queries to perform RAG\textsubscript{Dense}. This setting establishes an upper bound on retrieval effectiveness for RAG\textsubscript{Dense}, showing the potential of an ideal retrieval system.

\smallskip
The Callees and RAG\textsubscript{Dense-oracle} settings are unrealistic in practice, as developers cannot obtain the exact function dependencies beforehand nor have access to the correct solutions.
\subsection{Task Formulation}
Each data instance of \ours{} consists of a repository snapshot, target function signature, and docstring (from Section \ref{sec:benchmark_structure}). The task is to generate the target function that passes all relevant test cases. In practice, LLMs are provided with a prompt consisting of the retrieved context and asked to generate the solution.

Once the model generates the function, the synthesized code is inserted back into the repository to perform test execution. The solution is considered to pass if all relevant test cases pass. 
We provide the details of prompt construction in Appendix~\ref{sec:prompt_setup}.

We report the models' performance using Pass@1~\cite{codex} that measures the probability that the first generated code sample is correct.

\input{tables/eval_result}

\subsection{Model Setup}
Given the complexity of \ours{}, we evaluate only LLMs that meet the following criteria: state-of-the-art (SOTA) performance on existing benchmarks, and a context length of at least 16K.

Thus, we evaluate GPT-4o, GPT-4o-mini, DeepSeek-V2.5, and Claude 3.5 Sonnet, representing commercial LLMs. We also evaluate open-source LLMs such as CodeLlama, DeepSeek-Coder, and OpenCodeInterpreter series, ranging from 6.7B to 34B parameters. For commercial models, we use their official API, and for the open-source LLMs, we use the implementations provided by HuggingFace and vLLM~\cite{vllm}. Under each experimental setting (retrieval approach), we let each LLM generate one output per instance in \ours{} using greedy decoding.

%% file: tables/eval_result.tex
\begin{table}[t]
    \centering
    \scriptsize
    \setlength{\tabcolsep}{6pt}
    \begin{tabular}{lrrr}
        \toprule
        {\textbf{Models}} & {\textbf{RAG\textsubscript{BM25}}} &{\textbf{RAG\textsubscript{Dense}}} & {\textbf{Current-File}}  \\
        \midrule
CodeLlama-7B &\underline{10.7} &10.4 &5.7 \\
CodeLlama-34B &12.4 &\underline{12.8} &9.6 \\
DeepSeekCoder-6.7B &14.0 &\underline{14.1} &10.9 \\
DeepSeekCoder-33B &16.7 &\underline{17.1} &14.9 \\
OpenCodeInterpreter-6.7B &12.1 &{12.5} &\underline{13.2} \\
OpenCodeInterpreter-33B &15.3 &16.3 &\underline{18.3} \\
\midrule
Claude 3.5 Sonnet &14.4 &17.5 &\underline{19.8} \\
DeepSeek-V2.5 &18.5 &20.7 &\underline{27.0} \\
GPT-4o-Mini &15.1 &15.0 &\underline{18.7} \\
GPT-4o & \underline{27.4} &27.0 &26.8 \\
        \bottomrule
    \end{tabular}
    \caption{Pass@1(\%) of SOTA LLMs on \ours{}.}
    \label{tab:eval_result}
\end{table}

%% file: sections/Result.tex
\section{Result}
\subsection{SOTA LLMs' Pass@1 on \ours{}}
Table~\ref{tab:eval_result} shows ten LLMs' performance on \ours, under three retrieval approaches. On all retrieval approaches, commercial LLMs perform better. Specifically, GPT-4o has the best result, reaching up to 27.4\% pass@1. On the other hand, none of the open-sourced LLMs has over 20\% pass@1.%

This result shows that SOTA LLMs still struggle with repository-level code generation. Compared to their pass@1 on HumanEval (about 90\%~\cite{deepseekcoder}), SOTA LLMs are still \textbf{far away from writing real-world programs requiring repository-level information}. We provide a detailed discussion of the results below.

\newsubsubsection{Impact of Model Size}
Table~\ref{tab:eval_result} demonstrates the consistent advantage of larger models in solving complex repository-level tasks compared to smaller models within the same architecture. 

\begin{figure}[t]
    \centering
    \includegraphics[width=0.6\linewidth]{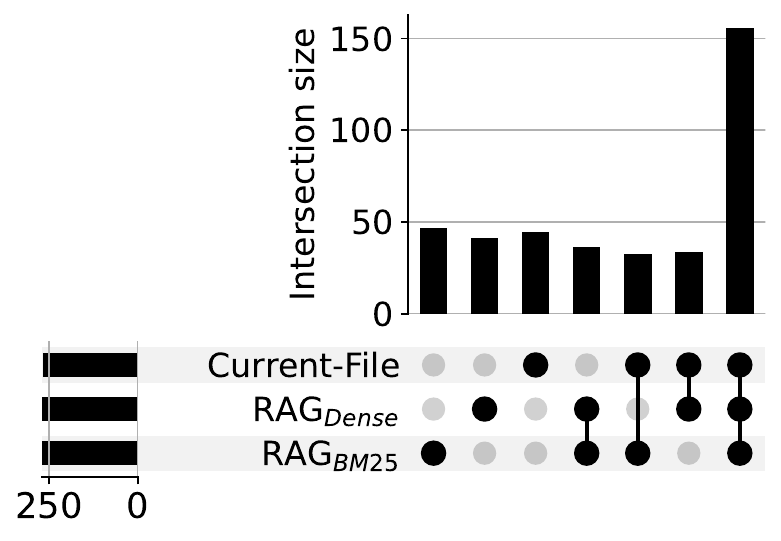}
    \caption{GPT-4o's passed tasks in different retrievals.}
    \label{fig:upset2}
\end{figure}

\begin{figure}[t]
    \centering
    \includegraphics[width=0.9\linewidth]{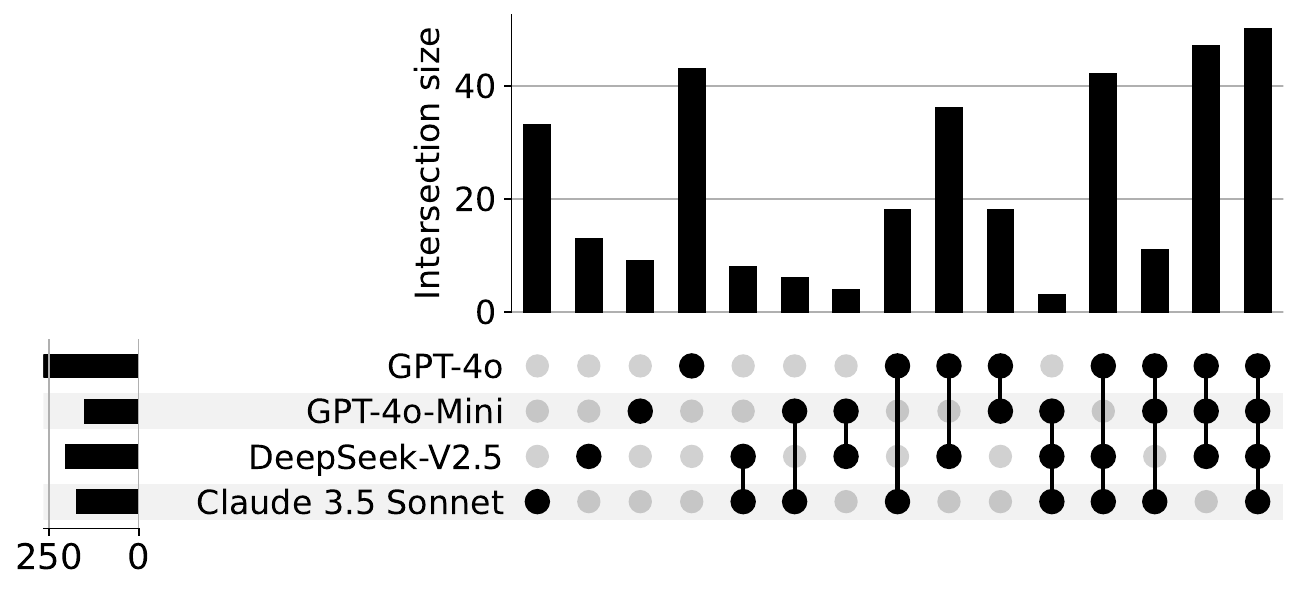}
    \caption{Commercial LLMs' passed tasks.}
    \label{fig:upset1}
\end{figure}

\newsubsubsection{Impact of Retrieval Approach}
The pass@1 results are higher for commercial models when using contexts from the current-file setting, while RAG\textsubscript{BM25} and RAG\textsubscript{Dense} yield similar but lower outcomes. This trend is also observed in the OpenCodeInterpreter series of LLMs. In contrast, other open-source LLMs perform better with RAG\textsubscript{BM25} and RAG\textsubscript{Dense} compared to the current-file setting. These results suggest that commercial models are more likely to benefit from contexts drawn from the current-file setting.

To further investigate the overlap and uniqueness of tasks solved using different retrieval approaches, %
Figure~\ref{fig:upset2} is an UpSet plot~\cite{upset} that illustrates the overlap and uniqueness of model results using GPT-4o's result. Though most solutions are solvable by all approaches, each approach has uniquely solved tasks. This suggests that retrieval approaches differ in overall effectiveness and may capture complementary contextual information.

\newsubsubsection{Model-Specific Solution Diversity}
Beyond retrieval approaches, we also analyze the distribution of uniquely solved tasks by each model using the same context.
Figure~\ref{fig:upset1} shows the number of passed tasks for each commercial LLM and their intersections in \ours under the RAG\textsubscript{Dense}. Notably, each model solves a unique set of tasks, highlighting the specialized capabilities of individual models. 
Particularly, Claude 3.5 Sonnet and GPT-4o each have a high number of uniquely solved tasks, indicating that they each excel at certain tasks.

\begin{figure*}[tb]
    \centering
    \begin{subfigure}[b]{0.32\linewidth} %
        \centering
        \includegraphics[width=\linewidth]{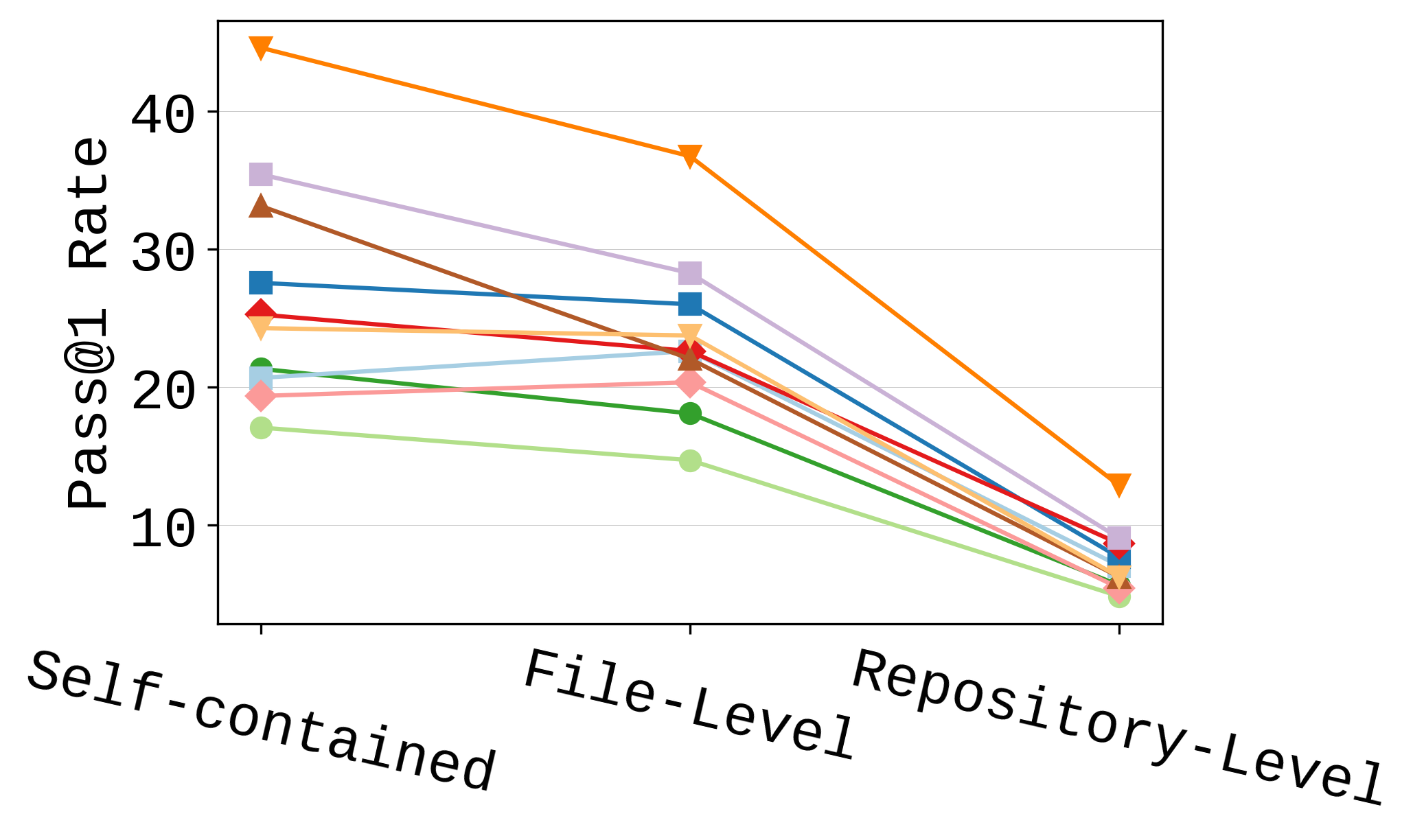} %
        \caption{Pass@1(\%) of LLMs across different context complexities.}
        \label{fig:complexity1}
    \end{subfigure}
    \begin{subfigure}[b]{0.32\linewidth} %
        \centering
        \includegraphics[width=\linewidth]{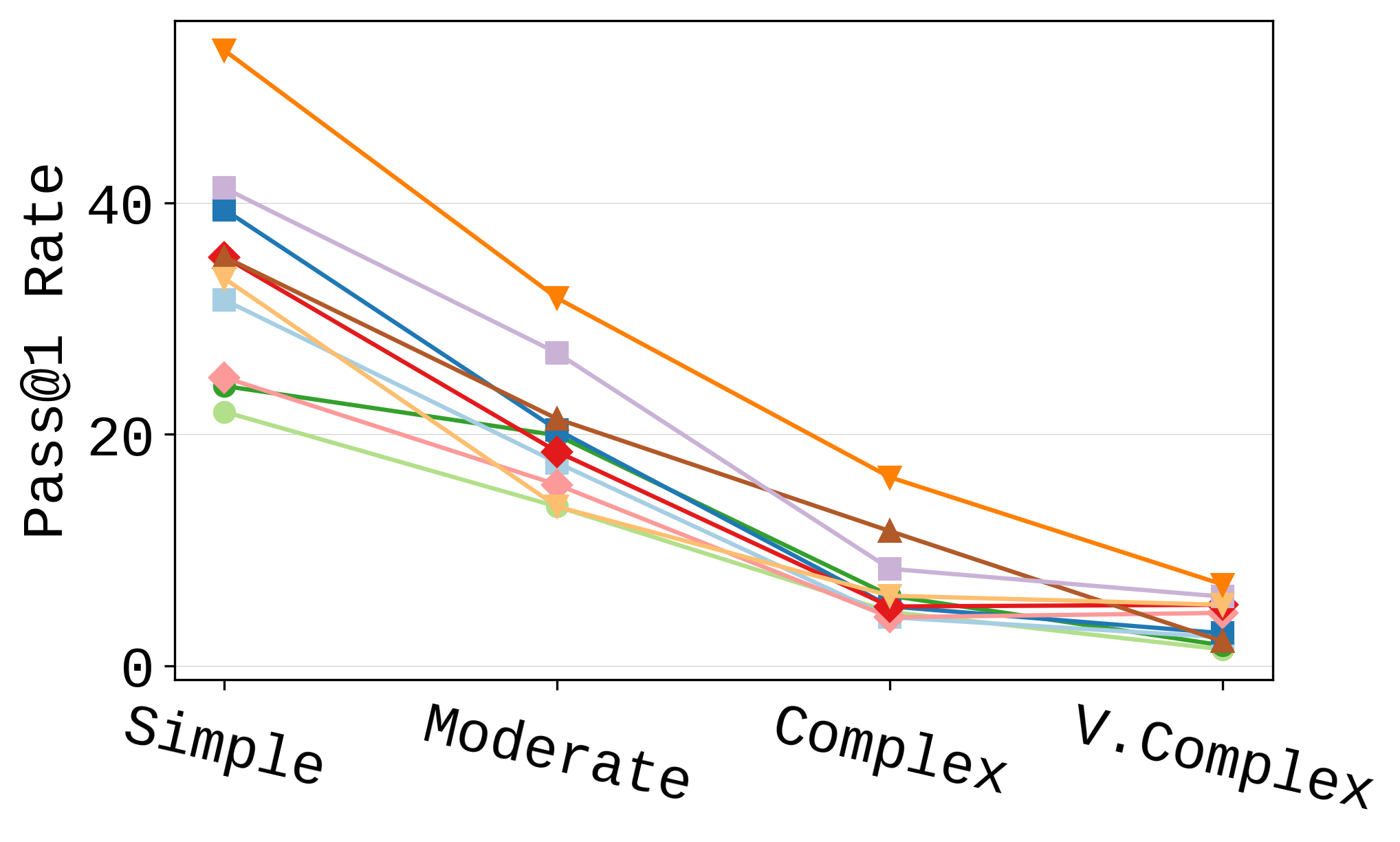} %
        \caption{Pass@1(\%) of LLMs across different cyclomatic complexities.}
        \label{fig:complexity2}
    \end{subfigure}
    \begin{subfigure}[b]{0.32\linewidth} %
        \centering
        \includegraphics[width=\linewidth]{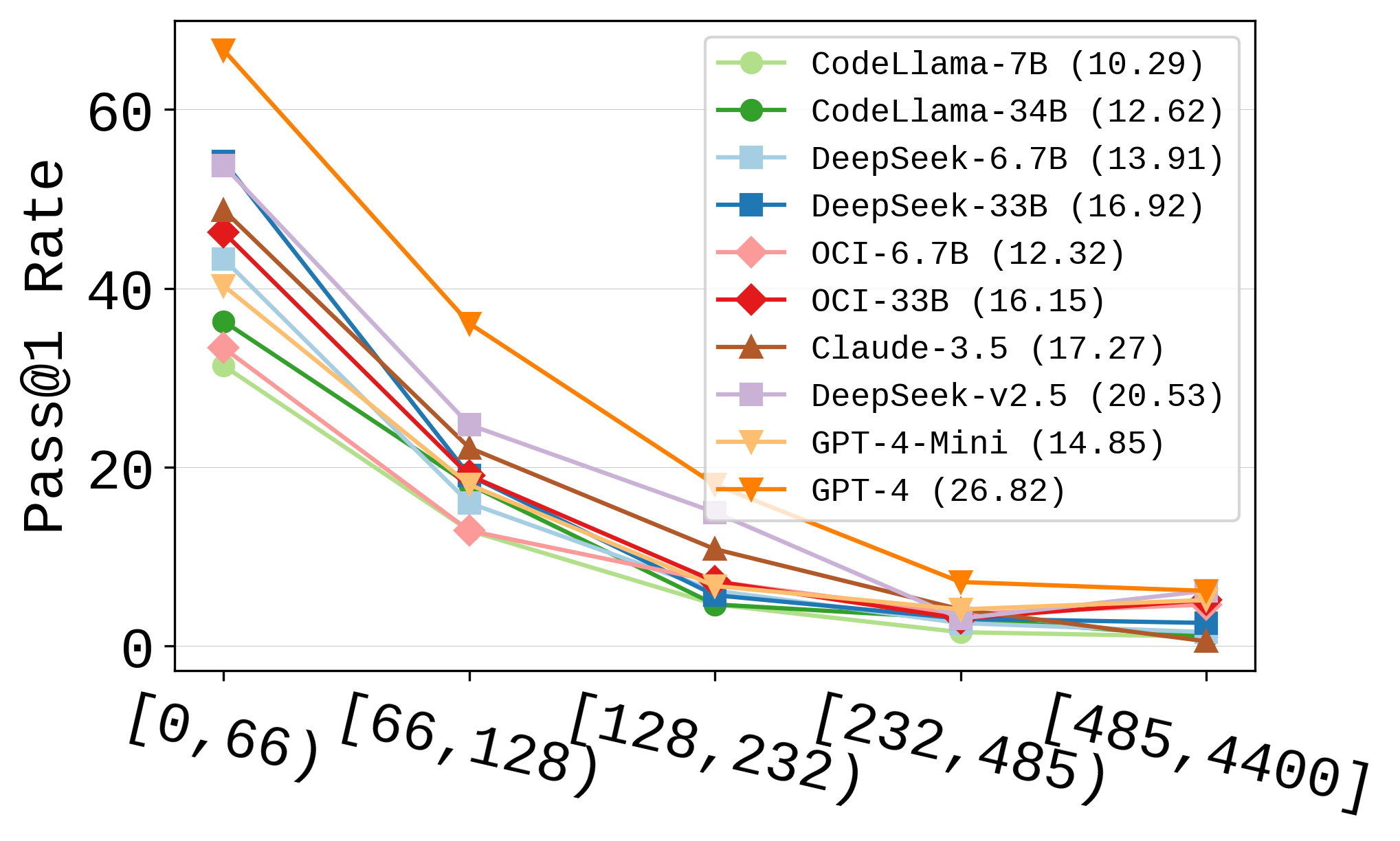} %
        \caption{Pass@1(\%) of LLMs across different length complexities.}
        \label{fig:complexity3}
    \end{subfigure}

    \setlength{\belowcaptionskip}{-5pt}
    \caption{Pass@1 under context complexity and length complexity. Let $n$ represent the number of instances in each bin. Length complexities distribution:: \textbf{[0, 66)}:
    $n$=201,  
    \textbf{[66, 128)} : $n$=194,  
    \textbf{[128, 232)}: $n$=194,  
    \textbf{[232, 435)}: $n$=196,  
    \textbf{[435, 4308]}: $n$=195.
    Cyclomatic complexity distribution ($M$):
\textbf{Simple} (\(M \leq 2\)):  $n$=269, 
\textbf{Moderate} (\(3 \leq M \leq 5\)): $n$=211 , 
\textbf{Complex} (\(6 \leq M \leq 10\)): $n$=215, 
and \textbf{Very Complex} (\(M \geq 11\)): $n$=285.}
\end{figure*}

\subsection{Impact of Complexity on Performance}
\label{subsec:complexity}
We show the performance of SOTA LLMs negatively correlates with the level of complexities. We compare three types of complexities: \textit{Context Complexity}, \textit{Canonical Solution Length}, and \textit{Cyclomatic Complexity levels}. We demonstrate the result in the RAG\textsubscript{Dense} setting.

\newsubsubsection{Context Complexity}
Figure~\ref{fig:complexity1} details the comparison of the evaluation results of SOTA LLMs for tasks with different context complexities. All models have the lowest pass@1 when generating functions with repository-level context compared to functions with less complex dependencies. The overall pass rate decreases as the complexity of the context level increases. 

\newsubsubsection{Cyclomatic Complexity}
Figure~\ref{fig:complexity2} shows that LLM performance declines as cyclomatic complexity increases. In the most challenging setting, the best LLM achieves only a 7.0\% pass@1 rate.

\newsubsubsection{Token Length}
Figure~\ref{fig:complexity2} presents the performance of LLMs on tasks that require generating functions of varying token lengths. The pass@1 of the LLMs gradually decreases as the length complexity of the functions increases. In the more challenging scenarios (length > 232), even the top-performing model, GPT-4o, achieves less than 10\% pass@1. %

With the highest level of complexities in all three categories, \ours establishes a new standard for LLMs in code generation.

\input{tables/recall_vs_pass_rate}

\subsection{Impact of Recall on Performance}
\label{sec:recall}
We study the impact of retrieval recall (recall of target function dependencies) on LLMs' pass rates in RAG\textsubscript{BM25} and RAG\textsubscript{Dense}. Table~\ref{tab:recall-vs-pass-rate} presents the pass@1 performance of LLMs for retrieved content across different recall rate ranges: 0, (0, 0.5], and (0.5, 1]. Interestingly, When recall falls within (0, 0.5], model performance is comparable to that with a 0 recall rate. In contrast,
\textbf{high recall (in range (0.5,1]) in retrieved content consistently leads to better model performance,} indicating the importance of retrieving target function dependencies.

\input{tables/oracle_retrieval}

\subsection{Impact of Context Type on Performance}
We compare the three retrieval approaches with three additional settings using GPT-4o and GPT-4o-mini in Table~\ref{tab:pass-rate-vs-context-complexity}. %

\textbf{Finding 1: SOTA LLMs underperform on \ours even with RAG\textsubscript{Dense-oracle}.} Despite using canonical solutions as queries, RAG\textsubscript{Dense-oracle} only marginally outperforms RAG\textsubscript{Dense}. Further analysis reveals that, on average, 22.4 of the top 30 retrieved functions \textbf{(75\%)} overlap between \colorbox{yellow!20}{\raisebox{0pt}[5pt][0pt]{RAG\textsubscript{Dense}}} and \colorbox{yellow!20}{\raisebox{0pt}[5pt][0pt]{RAG\textsubscript{Dense-oracle}}}, suggesting that task descriptions alone enable effective retrieval in \ours. This result underscores that even with an ideal retrieval setup, SOTA LLMs struggle on \ours{} tasks, highlighting the limitations of existing LLMs for real-world coding tasks.

\textbf{Finding 2: Additional context significantly improves performance for all types of tasks in \ours.}
All settings are better than the \colorbox{gray!20}{Baseline} setting. Even for self-contained tasks, the inclusion of 1,024 prefix tokens (Callees setting) improves the pass@1 result, suggesting that additional context—beyond just the task description—improves generation performance.

\textbf{Finding 3: Dependency-based context is suboptimal.}
The Callees setting is comparable to RAG\textsubscript{BM25} and RAG\textsubscript{Dense} but does not consistently outperform them. Across all complexity levels, it lags behind the Current-File setting. This indicates that solely using dependencies is not ideal for repository-level code generation. Combined with the findings from Section~\ref{sec:recall}, future work should aim for retrieving contents with a high recall of the target function's dependencies while maintaining high similarity to the task description.

\subsection{Failure Reasons}
We compute the ratio of syntax errors as the proportion of tasks with syntax errors among all failed samples and present the results of all combinations of models (10 models) and retrieval approaches in Figure~\ref{fig:syntax}.
 Results show that the Current-File setting leads to a higher syntax error ratio, and models with a higher syntax error ratio tend to achieve lower Pass@1 performance. For RAG\textsubscript{BM25} and RAG\textsubscript{Dense} methods, the syntax error ratio remains below 25\%, suggesting that LLMs are less likely to produce syntax errors when leveraging these retrieval strategies. %

We study GPT-4o's failures on 30 randomly sampled tasks and find two primary failure reasons. 
First, LLM-generated functions frequently lack sufficient input validation and parameter handling, such as ignoring or improperly transforming input values, leading to errors when processing unexpected inputs or misusing function arguments.
Second, incorrect or incomplete implementation of core logic leads to deviations from the intended functionality, including selecting suboptimal computational methods, mismanaging object states, or misapplying parallel processing strategies. %
We provide case studies in the Appendix~\ref{sec:failure analysis}.

\input{tables/wrong_syntax_percentage}

%% file: tables/recall_vs_pass_rate.tex
\begin{table}[t]
    \centering
    \scriptsize
    \setlength{\tabcolsep}{8pt}
    \begin{tabular}{l c r r r}
        \toprule
        \multirow{3}{*}{\textbf{Model}} & \multirow{3}{*}{\textbf{Method}} & \multicolumn{3}{c}{\textbf{Recall Range}} \\
        \cmidrule{3-5}
        &  & 0 & (0, 0.50] & (0.50, 1] \\
        \midrule
\multirow{2}{*}{DeepSeek-v2.5}  
        & RAG\textsubscript{BM25} & 10.4 & 10.9 & 28.8 \\
        & RAG\textsubscript{Dense} & 12.3 & 11.0 & 27.2 \\
        \midrule
\multirow{2}{*}{Claude-3.5}  
        & RAG\textsubscript{BM25} & 7.1 & 4.5 & 24.7 \\
        & RAG\textsubscript{Dense} & 7.2 & 8.7 & 28.3 \\
        \midrule
\multirow{2}{*}{GPT-4o}  
        & RAG\textsubscript{BM25} & 20.3 & 16.4 & 41.1 \\
        & RAG\textsubscript{Dense} & 16.7 & 14.2 & 38.0 \\
        \midrule
\multirow{2}{*}{GPT-4o-Mini}  
        & RAG\textsubscript{BM25} & 10.8 & 7.3 & 24.7 \\
        & RAG\textsubscript{Dense} & 7.7 & 9.4 & 28.3 \\
        \bottomrule
    \end{tabular}
    \caption{Pass@1 by Recall for RAG\textsubscript{BM25} and RAG\textsubscript{Dense}. Distribution: (BM25: 492/110/73, Dense: 456/127/92).}
    \label{tab:recall-vs-pass-rate}
\end{table}

%% file: tables/oracle_retrieval.tex
\begin{table}[t]
    \centering
    \scriptsize
    \setlength{\tabcolsep}{5.5pt}
    \begin{tabular}{l l r r r r}
    \toprule
    \textbf{Model} &\textbf{Method} & \textbf{Self} & \textbf{File} & \textbf{Repo.} & \textbf{Overall} \\
    \midrule
    \multirow{7}{*}{GPT-4o-Mini}
        & \cellcolor{gray!20} Baseline & \cellcolor{gray!20} 11.5 & \cellcolor{gray!20} 7.3 & \cellcolor{gray!20} 2.6 & \cellcolor{gray!20} 6.2 \\
        & \;RAG\textsubscript{BM25} & 22.6 & 22.0 & 8.0 & 15.1 \\
        & \cellcolor{yellow!20} RAG\textsubscript{Dense} & \cellcolor{yellow!20} 24.3 & \cellcolor{yellow!20} 23.7 & \cellcolor{yellow!20} 6.2 & \cellcolor{yellow!20} 15.0 \\
        & \;Current-File & 30.5 & 23.7 & 9.6 & 18.7 \\
    \cmidrule{2-6}
        & \;Callees & 27.9 & 23.7 & 7.0 & 16.5 \\
        & \cellcolor{yellow!20} RAG\textsubscript{Dense-oracle}  & \cellcolor{yellow!20} 28.2 & \cellcolor{yellow!20} 24.3 & \cellcolor{yellow!20} 5.2 & \cellcolor{yellow!20} 15.8 \\
    \midrule    
    \multirow{7}{*}{GPT-4o}
        & \cellcolor{gray!20} Baseline & \cellcolor{gray!20} 23.6 & \cellcolor{gray!20} 11.3 & \cellcolor{gray!20} 3.8 & \cellcolor{gray!20} 11.3 \\
        & \;RAG\textsubscript{BM25}  & 39.3 & 31.1 & 18.7 & 27.3 \\
        & \cellcolor{yellow!20} RAG\textsubscript{Dense} & \cellcolor{yellow!20} 44.6 & \cellcolor{yellow!20} 36.7 & \cellcolor{yellow!20} 12.9 & \cellcolor{yellow!20} 27.0 \\
        & \;Current-File & 39.3 & 35.0 & 16.3 & 26.8 \\
   \cmidrule{2-6}
        & \;Callees & 35.1 & 31.1 & 12.2 & 22.8 \\
        & \cellcolor{yellow!20} RAG\textsubscript{Dense-oracle} & \cellcolor{yellow!20} 45.2 & \cellcolor{yellow!20} 34.5 & \cellcolor{yellow!20} 16.3 & \cellcolor{yellow!20} 28.6 \\

    \bottomrule
    \end{tabular}
    \caption{Pass@1 (\%) of GPT-4o across context complexities. Self, File, and Repo. refer to Self-Contained, File-Level and Repository-Level.}
    \label{tab:pass-rate-vs-context-complexity}
\end{table}

%% file: tables/wrong_syntax_percentage.tex
\begin{figure}
    \centering
    \includegraphics[width=0.9\linewidth]{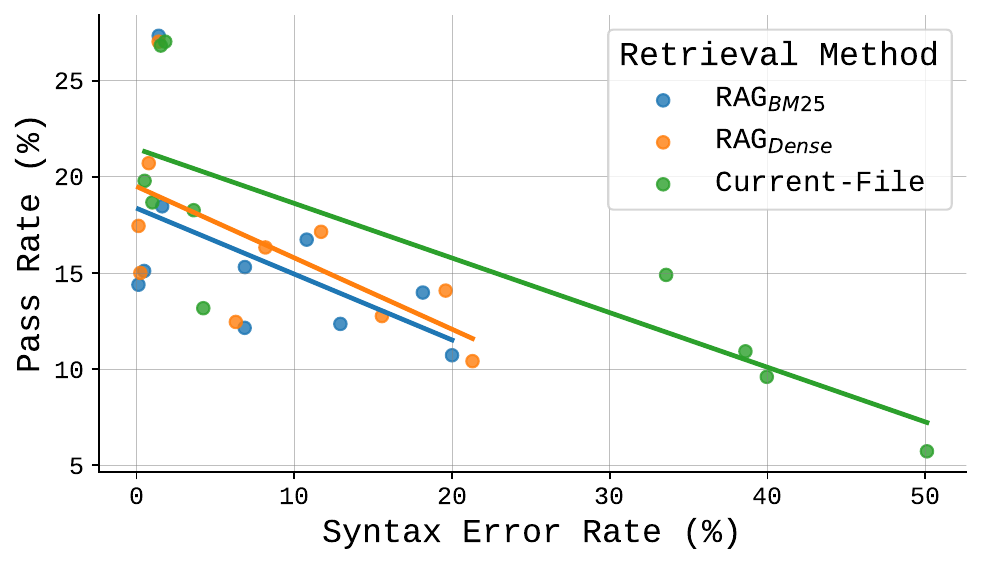}
    \caption{Models' pass@1 against syntax error rate in different retrievals. Each pair of model and retrieval approach is represented as a dot.}
    \label{fig:syntax}
\end{figure}

%% file: sections/Related_Work.tex
\section{Related Work}
\label{sec: related}

\subsection{LLMs for Code}
LLMs have been widely used for code-related tasks, among which code generation is the most important and common task~\cite{gpt4, starcoder, starcoder2, CodeGen, codellama, deepseekcoder, wizardcoder}. %
Existing work has shown an impressive ability to generate self-contained programs.

However, the performance of LLMs for code generation within the context of repositories remains underexplored. Despite the use of retrieval-augmented generation (RAG) to retrieve relevant context from the repository~\cite{crosscodeeval, codereval, repoformer}, generating code within the context of the repositories remains more challenging and yields lower accuracy compared to generating self-contained code.

\subsection{Code Generation Benchmarks}
Code generation benchmarks are important for evaluating LLMs' ability to generate code. Self-contained benchmarks~\cite{apps, alphacode, evoeval, PlusEval} are unable to evaluate LLMs' capability to generate code for real-world projects.

A few repository-level code generation benchmarks have been developed. Concode~\cite{related-concode} represents an early approach in this domain, focusing on repository-level code generation evaluated through similarity-based metrics rather than the test-based metrics used in \ours. CrossCodeEval~\cite{crosscodeeval}, RepoBench~\cite{repobench}, and Long-Code-Arena~\cite{longcodearena} are %
benchmarks collected from GitHub projects. They consist of tasks generating a single line of code given the incomplete code snippets and repository context. As discussed in the introduction, they share two limitations: (1) their target code is limited to single-line outputs, and (2) they rely on similarity-based evaluation metrics.

Other benchmarks, such as CoderEval~\cite{codereval}, RepoEval~\cite{related-repocoder-repoeval}, R2E-Eval1, EvoCodeBench~\cite{related-evocodebench} and DevEval\cite{realted-deveval} consists of function generation tasks in Python and/or Java tasks collected from GitHub repositories. However, they fall short in terms of task complexity and scale. For example, they utilize repositories with limited size, fewer than 243 files on average, significantly smaller than widely-used repositories such as scikit-learn (1,640 files by \ours's collected date).

There is also contemporary work such as SWE-Gym~\cite{related-swegym}, which also engages with code at the repository level. However, its primary focus is on training agents for issue fixing, distinguishing it from the code generation emphasis of the benchmarks previously discussed and the objective of \ours.

In contrast, \ours combines complex real-world tasks (analyzed in Section~\ref{sec: comparison_with_existing_benchmark}) from the most popular Python repositories with rigorous test-based assessments and consists of \samplenum code generation tasks that require LLMs to write large functions instead of single-line snippets, aligning model evaluation with the expectations of modern software development.

\newsubsubsection{Version-Aware Code Generation Benchmarks}
Recognizing the dynamic nature of software, another line of benchmarks specifically addresses API version evolution, a factor that can significantly influence code generation  performance. LibEvolutionEval~\cite{related-libevolutioneval} focuses on version-specific in-line code completion and documentation retrieval across library versions. GitChameleon~\cite{related-gitchameleon} provides manually curated Python problems with executable unit tests, conditioned on specific library versions to assess functional accuracy. CodeUpdateArena~\cite{related-codeupdatearena} uses synthetic API updates to evaluate knowledge editing in LLMs, testing if models can apply updates without in-context documentation. VersiCode~\cite{related-versicode} introduces tasks where models must complete code for a specific library version or update code to a different version. It uses a large, diverse dataset and its novel 'Critical Diff Check' metric, designed to specifically assess correct API usage across versions.

While the evolving nature of library versions and its impact on code generation is an important consideration, REPOCOD and the aforementioned version-focused benchmarks evaluate different facets of this challenge. They primarily target the effect of package versions on LLMs' code generation ability, often through simpler tasks such as token/line level completions, specific API updates, or less complex function generation. In contrast, REPOCOD evaluates LLMs' ability to solve complex tasks requiring a deep understanding of repository-level context. 

\subsection{Automated Test Collection Pipelines}
Various approaches have been proposed for collecting benchmarks with repository-level task instances~\cite{codereval, related-repocoder-repoeval, related-r2e, realted-deveval}. RepoEval executes all tests in the repositories but is impractical for popular repositories due to the sheer volume of test cases. R2E-Eval1, on the other hand, uses LLMs to generate equivalence tests. However, this strategy lacks verification of the correctness of the test cases, limiting its reliability for trustworthy evaluations.

The most comparable approach to ours is proposed in CoderEval, which employs static analysis to identify test cases that reach the target function and supplements them with manually curated test cases. While this approach works for small repositories, it is neither scalable—since it requires human effort—nor does it fully utilize the test cases provided by developers, as static analysis cannot capture all dependencies.

In contrast, our approach uses dynamic analysis to automatically identify test cases from existing developer tests. This execution-based approach is more scalable and is better suited for leveraging existing developer-provided test cases in popular repositories. Additional details on our data collection pipeline are provided in Section~\ref{sec:data_collection}.

%% file: sections/Conclusion.tex
\section{Conclusion}
We present \ours{}, a real-world, complex dataset designed for code generation tasks with repository-level context. \ours{} comprises \samplenum{} instances from 11 popular Python repositories, including \crossfilenum{} that require repository-level context, with canonical solutions averaging 331 tokens in length—highlighting the benchmark's complexity and comprehensiveness. In addition, We introduce a scalable automatic extraction method for collecting repository-level code generation tasks. Our evaluation of SOTA LLMs on \ours{} reveals a maximum pass@1 of 28.6\%, with even lower scores for functions requiring repository-level context, showing that existing LLMs fall short of generating realistic repository-level code. This work underscores the need for further research in repository-level code generation.

%% file: sections/Limitation.tex
\section{Limitation}
Our work has limitations. First, we collect data from only 11 repositories, covering a small subset of those that could be included in this benchmark; future versions of \ours{} will expand to other popular Python repositories. Second, we evaluate only ten models, representing a subset of popular LLMs. With more time and resources, we could test a broader range, providing a more comprehensive view of LLM capabilities in repository-level code generation. We will publish \ours{} so the community can evaluate additional models, broadening the scope of tested LLMs.

%% file: sections/Acknowledgement.tex
\section{Acknowledgements}
We thank the anonymous reviewers for their feedback on
this work. This research was supported in part by NSF 1901242 and 2006688 and a CFI fund.

%% file: sections/Appendix.tex
\clearpage 
\appendix
\section{Appendix}
\subsection{Basic Statistics of \ours}
\input{tables/ours_distributed_context_complexity}

\ours contains \samplenum  code generation tasks  from 
 \projectnum widely-used repositories, covering a wide range of functionalities, including data science, scientific computing, web development, and software development tools.

Table~\ref{tab:ours_complexity} presents detailed statistics for instances from each repository, categorized by context complexity types: \textit{repository-level}, \textit{file-level}, and \textit{self-contained}. For each category, it shows \textbf{\#NL} (number of tokens in target function descriptions), \textbf{\#GT} (number of tokens in canonical solutions), \textbf{Cyclo.} (average cyclomatic complexity of the canonical solution)~\cite{cyclomatic}), and \textbf{\#Funcs.} (number of target functions). Additionally, we report repository statistics: \textbf{\#Line} (average lines in Python files per repository) and \textbf{\#Files} (average Python files per repository).

We define three types of context complexities: \textit{Repository-level} involves canonical solutions that call functions from other files in the repository; \textit{file-level} involves calls to functions within the same file (excluding the target function) but not from other files; and \textit{self-contained} includes functions that only use commonly used libraries (e.g., \code{numpy}) or Python built-in modules (e.g., \code{os}).

Among the three settings, the repository-level functions have the longest token length for the canonical solutions (441.2), compared to file-level functions (192.0) and self-contained functions (233.8). Additionally, the repository-level functions have the highest cyclomatic complexity (10.8) compare to other two categories. The highest length and cyclomatic complexity with the addditional repository-level context makes repository-level the most challenging category. 

\subsection{Motivating Example}
\begin{figure}[h]
    \centering
    \includegraphics[width=\linewidth]{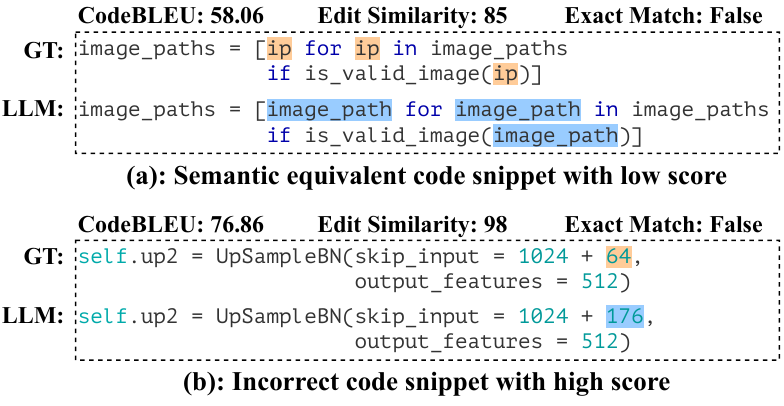}
    \caption{Two examples from RepoBench showing misleading metrics results. The yellow and blue highlights indicate the difference between ground truth (GT) and the LLM-generated code.}
    \label{fig: motivating}
\end{figure}
\begin{figure*}[ht]
    \centering
    \includegraphics[width=\linewidth]{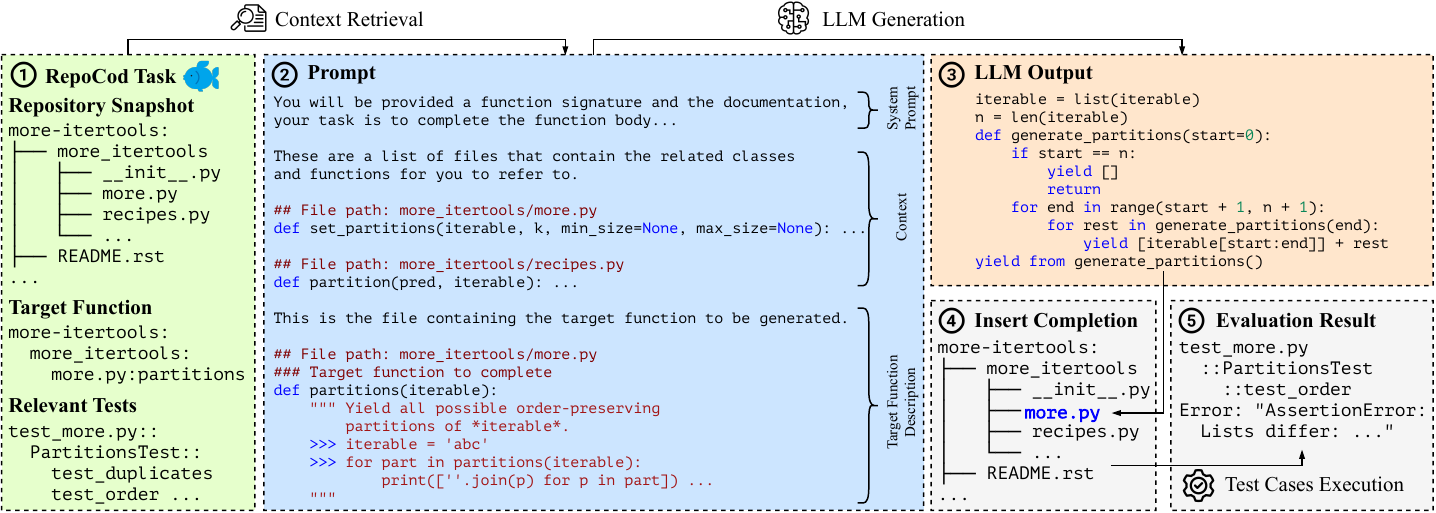}
    \caption{Illustrative example of prompting LLMs and evaluating the LLM-generated code.}
    \label{fig:illustrating_example}
\end{figure*}

\label{sec:motivating-example}

Figure~\ref{fig: motivating} (a) shows two RepoBench examples where LLM-generated code is semantically correct but wrongly penalized by CodeBLEU, edit similarity, and exact match. In contrast, Figure~\ref{fig: motivating} (b) demonstrates how these metrics assign high scores to incorrect solutions, underscoring their limitations in accurately evaluating generated code.

\subsection{Prompting Format}
\label{sec:prompt_setup}
As described in Section~\ref{sec:benchmark_structure}, the data instance of \ours consists of the repository snapshot, target function signature, and docstring. Figure~\ref{fig:illustrating_example} demonstrates an example of the prompt construction process, the format of the prompt, and the evaluation pipeline used for all our experiments.

The section highlighted in green represents the data instance of \ours (step \raisebox{.5pt}{\textcircled{\raisebox{-.9pt} {1}}}). The format of the prompt is detailed in the section highlighted in blue (step \raisebox{.5pt}{\textcircled{\raisebox{-.9pt} {2}}}). Our prompt consists of the system prompt, the file path to the target function (relative to the repository root), the retrieval context, and the target function description (the function signature and docstring of the target function). The retrieval context contains the relative file path, as well as the signature, the docstring, and the body of the retrieved functions. If the context exceeds LLM's context window, we truncate it from the beginning. 

Once the LLM generates the solution (step \raisebox{.5pt}{\textcircled{\raisebox{-.9pt} {3}}}), the code snippet is inserted into the repository for testing (step \raisebox{.5pt}{\textcircled{\raisebox{-.9pt} {4}}}, modified file highlighted in blue). Then \ours executes the relevant test cases (step \raisebox{.5pt}{\textcircled{\raisebox{-.9pt} {5}}}) and determines the correctness of the solution (step \raisebox{.5pt}{\textcircled{\raisebox{-.9pt} {5}}}). As one of the test cases fails, this generated code snippet is considered incorrect.

\subsection{Evaluation Result By Repository}

Table~\ref{tab:eval_result_by_repo} presents detailed LLM performance on \ours, revealing several key insights.
First, repositories differ significantly in difficulty. Models perform well on \code{flask}, with pass@1 often above 40\% , indicating an alignment with LLM capabilities. Second, GPT-4o is the strongest LLM on \ours, consistently outperforming or matching other models across all repositories. Finally, LLMs exhibit specialization across repositories. For instance, while Claude 3.5 Sonnet and DeepSeekCoder-33B perform similarly overall, Claude excels on \code{pylint}, whereas DeepSeekCoder performs better on \code{plotly.py}.

\input{tables/eval_result_per_repo}

\subsection{Recall of Retrieval approach}

\input{tables/retrieval_stat_by_repo}
Table~\ref{tab:retrieval-stat-by-repo} presents the recall rates of two different retrieval settings—RAG\textsubscript{BM25} and RAG\textsubscript{Dense}—applied across all selected repositories. These repositories include \code{pylint}, \code{sphinx}, \code{seaborn}, \code{flask}, \code{sympy}, \code{more-itertools}, \code{scikit-learn}, \code{xarray}, \code{datasets}, \code{plotly.py}, and \code{astropy}. The recall rates reflect performance at both the repository-level and file-level contexts, providing insights into how each setting performs for individual repositories as well as overall.

The recall rates for RAG\textsubscript{Dense} retrieval are generally higher than those for RAG\textsubscript{BM25} retrieval across most repositories, highlighting the effectiveness of the RAG\textsubscript{Dense} method in retrieving the invoked functions relevant to the target function.

There are notable variations between repositories. For instance, \code{more-itertools} and \code{plotly.py} show significantly higher recall rates with the RAG\textsubscript{Dense} setting, while \code{scikit-learn} demonstrates relatively low recall rates in both settings.

The total recall rate is computed using micro-averaging, and the result shows that neither retrieval method achieves a high recall for the oracle contexts. 

\subsection{Inference Length V.S. GT length}
\input{tables/inference_len_vs_correct_rate_dense}
We analyze the relationship between inference length and the length of canonical solutions across all retrieval settings. Table~\ref{tab:inference-len-dense-unfiltered} reports the average inference-to-canonical solution length ratio using micro-averaging. \textbf{On average, LLM-generated results are much longer than canonical solutions.}

We further investigate the causes of excessive generation length and summarize the following contributing factors:

Firstly, it is common for those underperforming LLMs to \textbf{generate repetitive code snippets} until they reach the token limit.

Secondly, despite being instructed to generate only the target function, LLMs sometimes \textbf{produce additional functions}, resulting in longer outputs. One contributing factor is that when models reference functions not retrieved in the context, they generate the missing helper function definitions, further inflating the output length.

To address this, we apply a filtering method that retains only the first unnested function in the generated output. Specifically, we discard all generated tokens following the first complete function that matches the target function name. Table~\ref{tab:inference-len-filtered-dense} presents the average filtered inference-to-canonical solution length ratio using micro-averaging. We make two key observations:
(1) Failed generations consistently exhibit longer relative lengths after filtering, particularly in Current-File retrieval settings, suggesting that verbosity may correlate with incorrect solutions.
(2) LLMs consistently generate longer code snippets than human developers, even when only the first unnested function is considered, indicating a systematic tendency toward verbosity.

\begin{figure}[t]
    \centering
    \includegraphics[width=\linewidth]{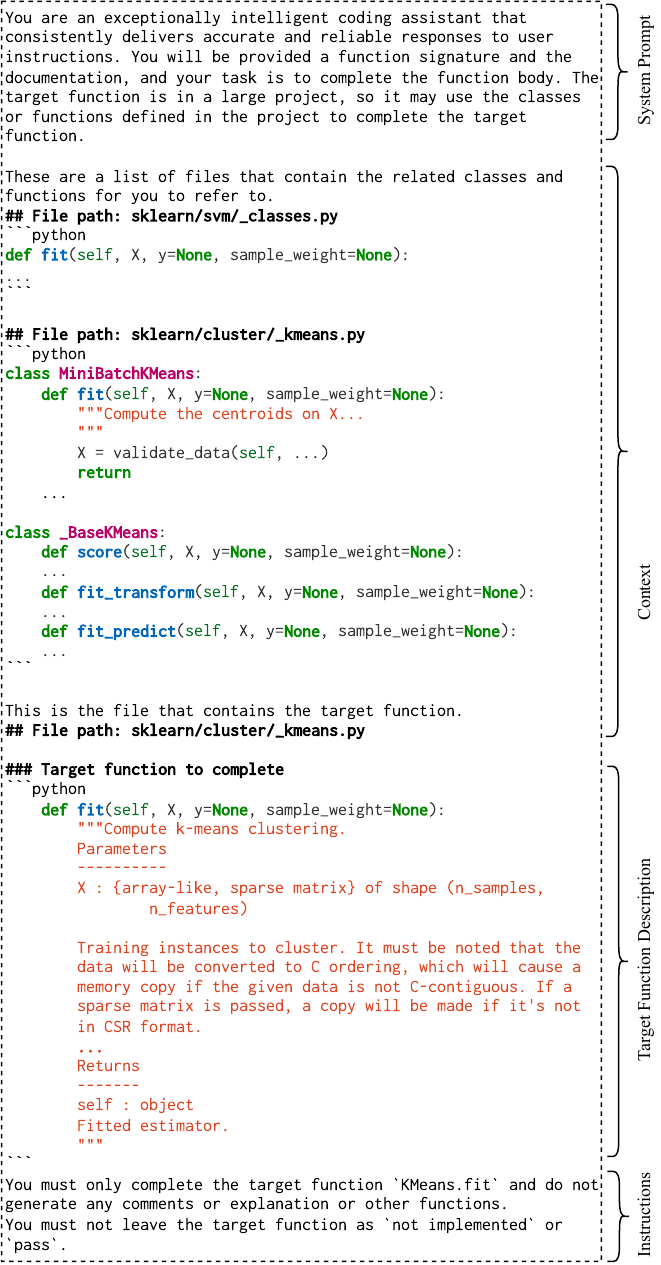}
    \caption{Prompt Example for \code{KMeans.fit} in the file ``\code{sklearn/cluster/\_kmeans.py}'' from \ours.}
    \label{fig:correct example}
\end{figure}

\newsubsubsection{Model-Specific Insights}

GPT-4o exhibits similar generation ratios before and after filtering, suggesting it adheres more closely to the task instructions and avoids unnecessary verbosity.

Claude-3.5 shows an inverse trend, where successful generations tend to be longer than failed ones (e.g., RAG\textsubscript{Dense}: Pass = 25.5 vs. Fail = 1.7). This suggests that Claude-3.5 generates and utilizes helper functions effectively, which may contribute to its success.

\input{tables/pass_vs_callee_len}
\subsubsection{Detailed context complexity's impact on Performance}
Table~\ref{tab:pass-rate-vs-callees-length} demonstrates the relationship between pass@1 and the detailed context complexity ( number of unique dependencies) present in the context. Across all models, performance declines as the number of dependencies increases, highlighting the growing challenge posed by more complex contextual requirements. For instance, DeepSeek-v2.5 with the BM25 retriever achieves a 31.8\% pass rate when no dependencies are involved, but this drops to 5.7\% when the number of dependencies reaches [5,35]. %

Notably, retrieval strategies significantly impact performance; models using the Current-File retrieval generally perform better for moderate dependency counts (e.g., DeepSeek-v2.5 achieving 17.3\% at [3,4]), suggesting that partial in-file context remains beneficial before complexity outweighs retrieval effectiveness. However, even strong retrieval strategies cannot fully mitigate the difficulty of handling a high number of dependencies, reinforcing the intuition that increasing context complexity directly correlates with reduced task success.

\subsection{Case Studies}

\begin{figure*}[t]
    \centering
    \includegraphics[width=\linewidth]{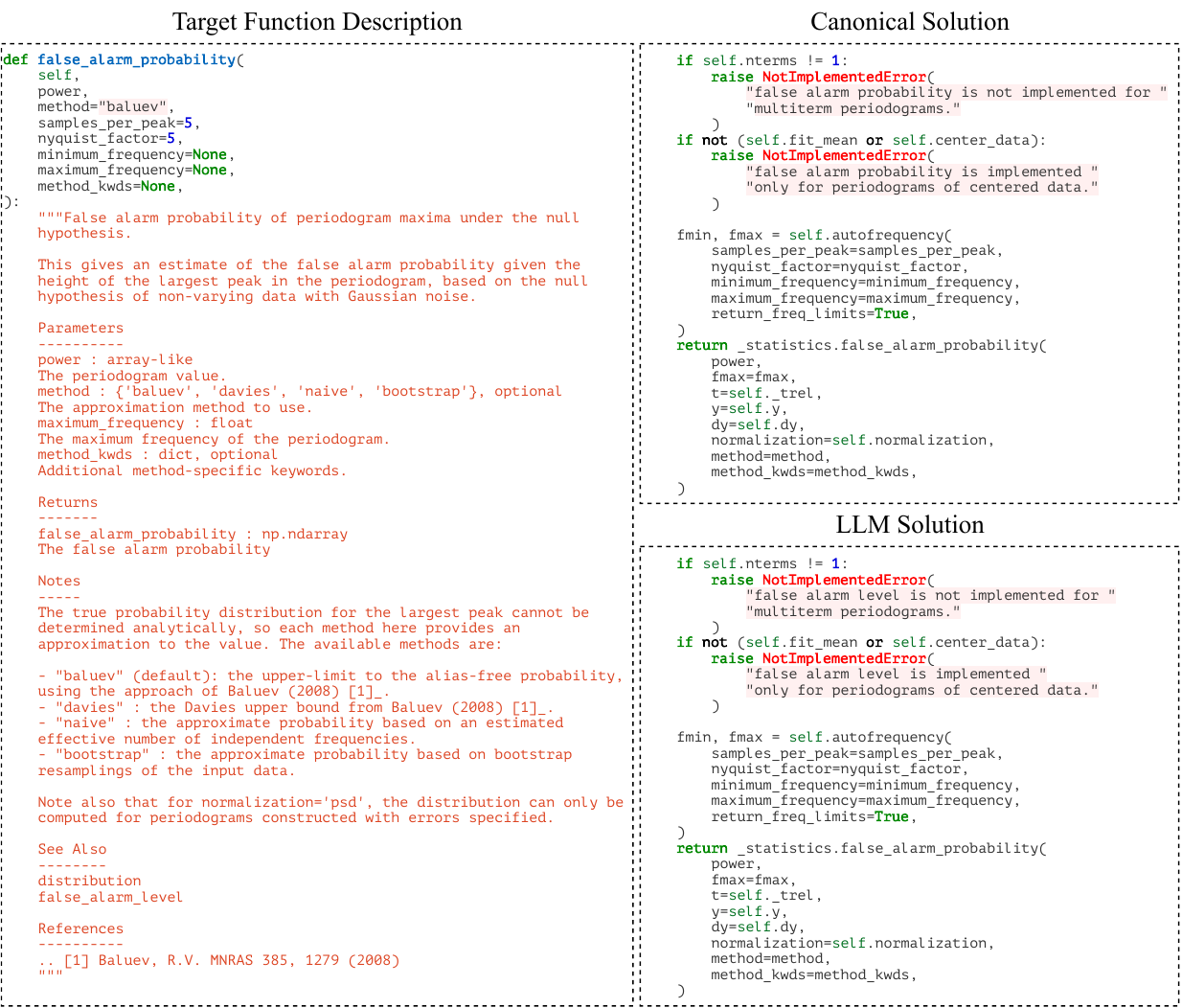}
    \caption{An example of the correct output generated by LLMs, for function ``\code{false\_alarm\_probability}'' in the file ``\code{astropy/timeseries/periodograms/lombscargle/core.py}'' from \ours{}.}
    \label{fig:correct_output}
\end{figure*}
\newsubsubsection{Prompt Example}
Figure~\ref{fig:correct example} shows an example of a \ours prompt, consisting of the system prompt, context information retrieved by RAG\textsubscript{BM25}, and the target function details, with specific content replaced by `\code{...}'.

\newsubsubsection{Passed Example}
Figure~\ref{fig:correct_output} shows an example of a correctly generated solution from \ours generated by GPT-4o under a RAG\textsubscript{Dense} retrieval setting. This example passes all test cases and is considered correct. Upon inspection, the generated code snippet is identical to the canonical solution except for the text content in the exception message.

\subsubsection{Failed Examples}

\label{sec:failure analysis}
\begin{figure*}[t]
    \centering
    \includegraphics[width=\linewidth]{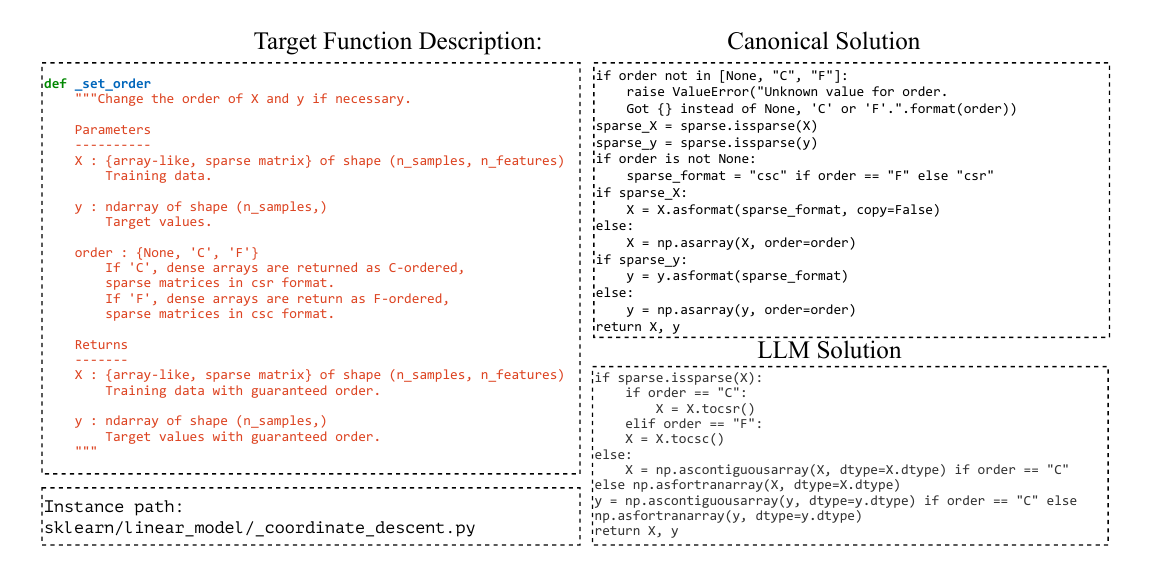}
    \caption{A failed case study, for function `\code{\_set\_order}' in the file `\code{sklearn/linear\_model/\_coordinate\_descent.py}' from \ours{}.}
    \label{fig:fail_case_study}
\end{figure*}

Figure \ref{fig:fail_case_study} shows an example of an incorrect solution generated by GPT-4o under RAG\textsubscript{BM25} retrieval setting. The target function \code{\_set\_order} is designed to change the order of training data (\code{X}) and target values (\code{y}). The canonical solution validates \code{order} is one of \code{[None, "C", "F"]}, raising a ValueError if an invalid value is provided. However, for the generated code, it does not validate \code{order}, which might cause unexpected behavior when \code{order} is incorrectly specified. 

In addition, the canonical solution maintains proper sparse matrix handling by explicitly checking \code{X} and \code{y} before conversion, while the generated code can lead to errors if \code{y} is sparse, since \code{np.ascontiguousarray()} and \code{np.asfortranarray()} do not support sparse matrices.

\begin{figure*}[t]
    \centering
    \includegraphics[width=\linewidth]{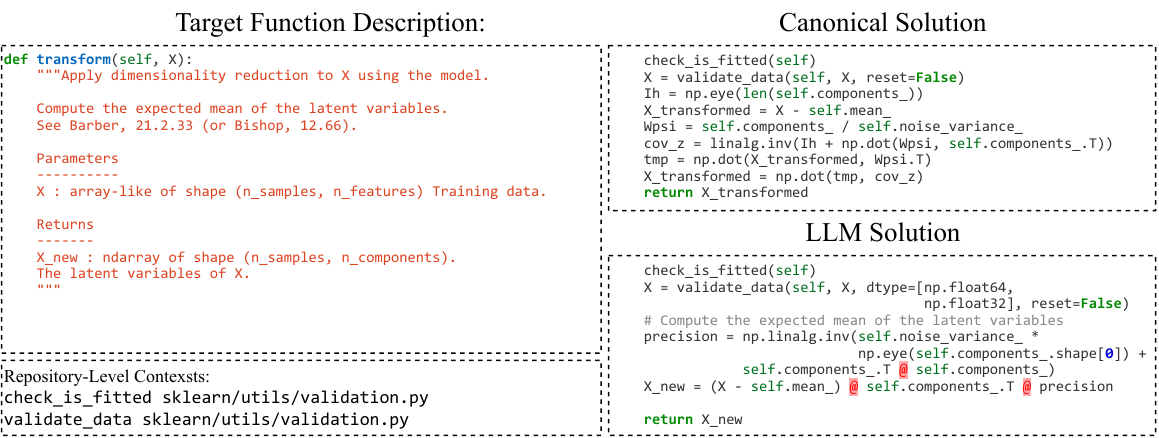}
    \caption{An example of the wrong output generated by LLMs, for function ``\code{transform}'' in the file ``\code{sklearn/decomposition/\_factor\_analysis.py}'' from \ours{}.}
    \label{fig:wrong_output}
\end{figure*}

Figure~\ref{fig:wrong_output} shows another example of an incorrectly generated solution by GPT-4o under RAG\textsubscript{BM25} retrieval setting. The canonical solution incorporates repository-level context; however, the solution generated by GPT-4o fails to match the functionality of the canonical solution. Specifically, the incorrect solution miscomputes the posterior mean by using an incorrect formulation for the covariance of the latent variables.

\begin{figure*}[tb]
    \centering
    \begin{subfigure}[b]{0.47\linewidth} %
        \centering
        \includegraphics[width=\linewidth]{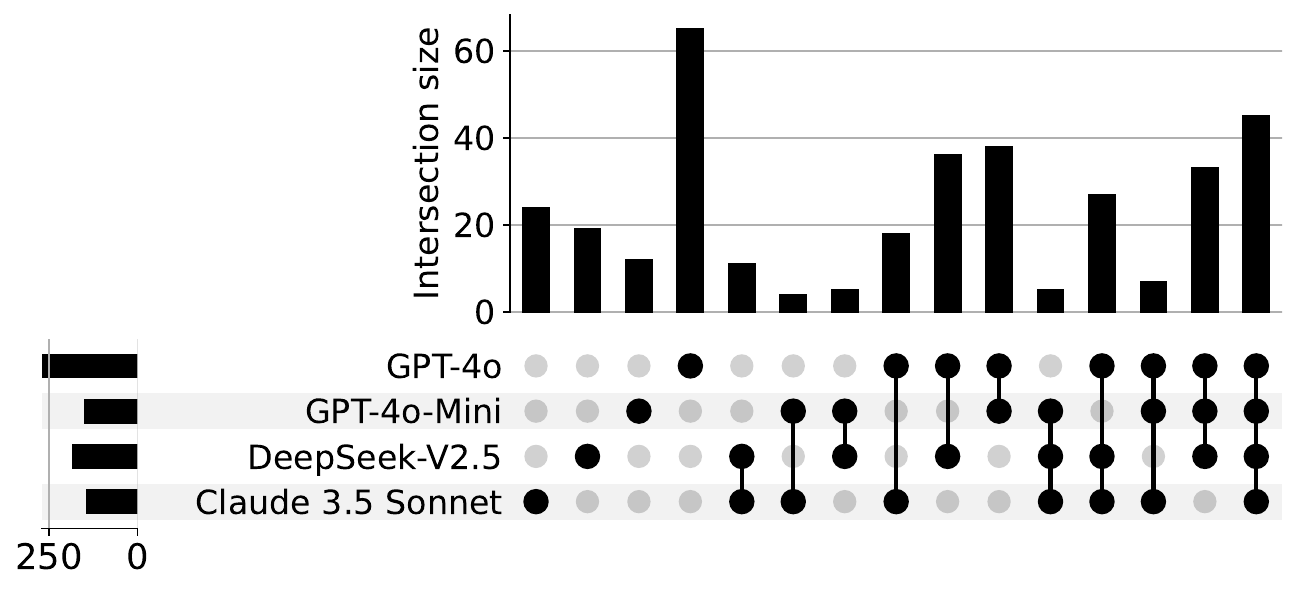} %
        \caption{Result for RAG\textsubscript{BM25}.}
        \label{fig:upset_commercial_bm25}
    \end{subfigure}
    \begin{subfigure}[b]{0.47\linewidth} %
        \centering
        \includegraphics[width=\linewidth]{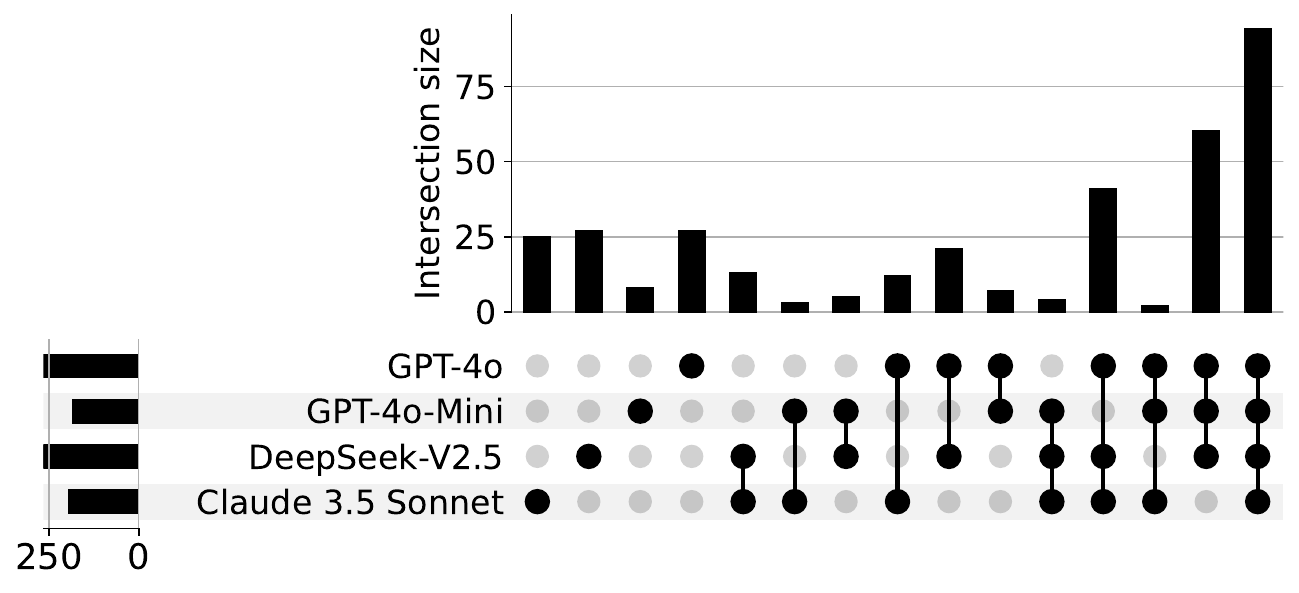} %
        \caption{Result for Current-File.}
        \label{fig:upset_commercial_current}
    \end{subfigure}
    \caption{Correct generation result's relationship under different retrieval settings.}
\end{figure*}

\subsection{Uniqueness and Overlap of Correct Generations}
\newsubsubsection{Commercial LLMs with RAG\textsubscript{BM25}}
Figure~\ref{fig:upset_commercial_bm25} presents the UpSet plot comparing the performance of four commercial models using contexts from RAG\textsubscript{BM25} retrieval, the largest overlap (more than 40) is shared by all models, indicating a significant degree of commonality. GPT-4o has the highest number of unique cases (over 60), suggesting its ability to capture distinct results in this scenario.

\newsubsubsection{Commercial LLMs with Current-File}
Figure~\ref{fig:upset_commercial_current} shows the Current File retrieval setting result. The overlap between all models increases greatly compared to the RAG\textsubscript{BM25} retrieval (to over 75 cases), demonstrating stronger alignment across models with this retrieval setting. In addition, the unique solvable problem by GPT-4o reduces greatly, to a similar level compared to Claude 3.5 Sonnet and DeepSeek-V2.5. 

\subsection{Impact of Retrieval Methods on Pass@1 across Varying Context Complexities}
\begin{figure*}[tb]
    \centering
    \begin{subfigure}[b]{0.49\linewidth} %
        \centering
        \includegraphics[width=\linewidth]{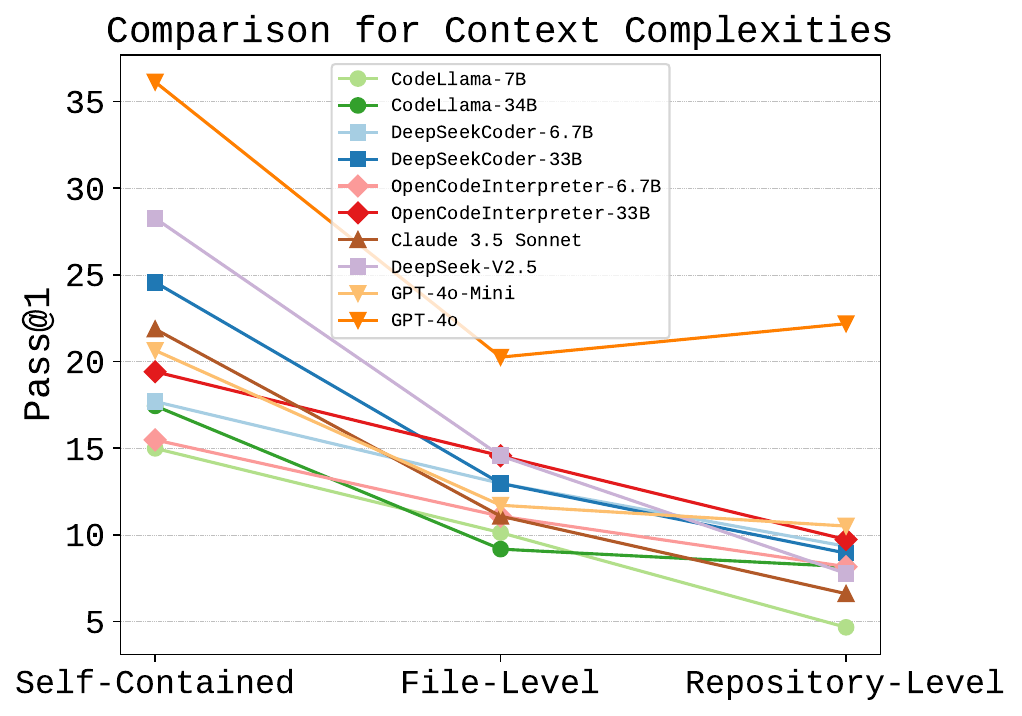} %
        \caption{Pass@1(\%) for LLMs based on RAG\textsubscript{BM25} retrieval.}
        \label{fig:context_complexity_bm25}
    \end{subfigure}
    \begin{subfigure}[b]{0.49\linewidth} %
        \centering
        \includegraphics[width=\linewidth]{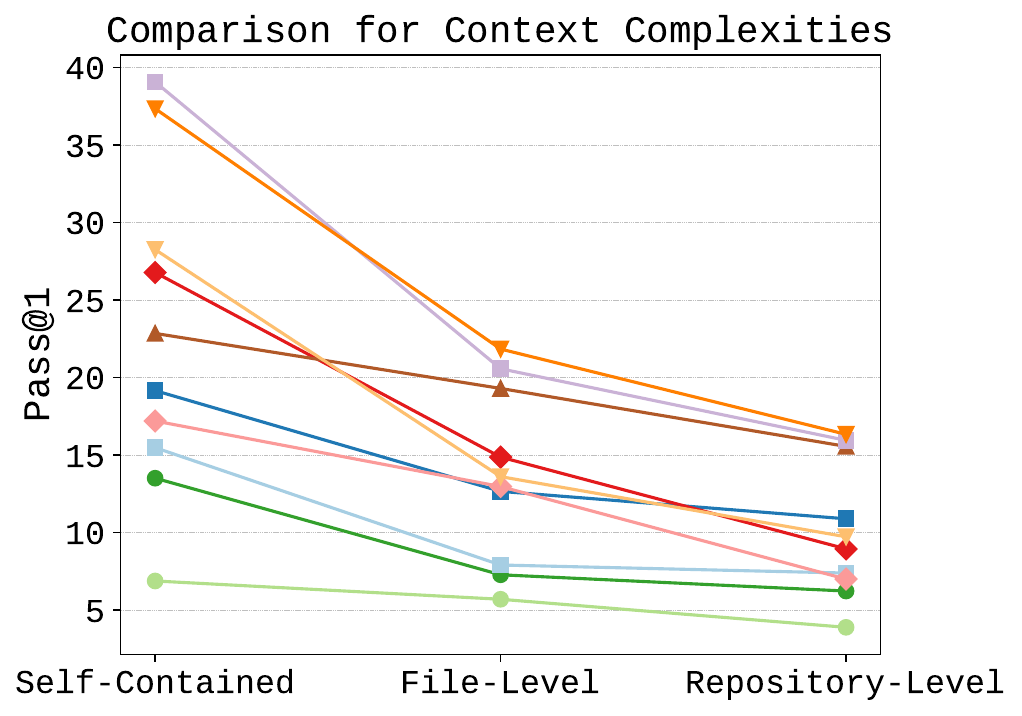} %
        \caption{Pass@1(\%) for LLMs based on current file retrieval.}
        \label{fig:context_complexity_current_file}
    \end{subfigure}
    \caption{Pass@1 under different context complexity.}
\end{figure*}

Figure~\ref{fig:context_complexity_bm25} and Figure~\ref{fig:context_complexity_current_file} illustrate the pass@1 for instances across various context complexities under different retrieval settings: RAG\textsubscript{BM25} and current file. In both settings, LLM performance tends to decrease as context complexity rises, progressing from self-contained to file-level and eventually to repository-level contexts. This trend aligns with the insights discussed in Section~\ref{subsec:complexity}.

An interesting finding is that GPT-4-o achieves a higher pass@1 rate for instances under repository-level context complexity compared to file-level instances. This observation suggests two potential insights: (1) specific retrieval methods, like RAG\textsubscript{BM25}, may enhance performance on repository-level task instances; (2) higher context complexity does not necessarily result in reduced LLM performance.

\subsection{\ours{-LITE}}

Given the large number of instances and the high evaluation cost of \ours, we introduce \ours{-LITE}—a 200 sample subset of the most challenging problems in \ours{—enabling faster and more targeted evaluation}. We define difficulty based on the product of the prompt length and canonical solution length (measured in lines of code), selecting samples with the highest scores. \ours{}-LITE maintains the original benchmark’s diversity by including problems from all three categories: self-contained, file-level, and repo-level, with 66, 67, and 67 samples, respectively. This subset enables efficient model evaluation while preserving the complexity of the coding tasks found in the full benchmark. We show the details of \ours{-LITE} in Table~\ref{tab:ours_complexity_by_repo_lite}. The performance of models on \ours{-LITE} is shown in Table~\ref{tab:lite-pass-rates}. The SOTA models' performance is consistently lower on \ours{-LITE} compared to that on \ours{}.

\input{tables/lite_result}

\subsection{OpenHands's Result on \ours{-LITE}}
To further explore the performance of advanced code generation approaches on \ours{} and address the applicability of our benchmark to code agents, we conducted additional experiments using OpenHands, a leading open-source agent framework. For this evaluation, we utilized \ours{}-LITE

In our experiments, OpenHands, when paired with Claude-3.5 Sonnet as its underlying model, successfully solved 11 out of 200 tasks (5.5\% Pass@1) in a single generation round. This result represents an improvement over the performance of Claude-3.5 Sonnet with standard Retrieval-Augmented Generation (RAG) approaches on \ours{}-LITE.

However, despite this improvement, OpenHands with Claude-3.5 Sonnet still underperformed the best RAG-based result (GPT-4o achieving 9.0\% Pass@1 with RAG$_{BM25}$).

These preliminary findings suggest that while current agentic approaches like OpenHands can offer incremental performance gains on complex, repository-level code generation tasks, they still face significant challenges on \ours{}. The difficulty of the \ours{} benchmark, underscores the substantial room for improvement in both LLM capabilities and agentic framework strategies for real-world code generation. We believe these evaluations provide a solid foundation and demonstrate \ours{}'s utility for benchmarking future advancements in code agents.

\input{tables/ous_lite_distributed_context_complexity}

\subsection{Potential Risk and Impact}
This work aims to construct an evaluation 
benchmark for code generation using real-world, 
repository-level context. We do not foresee any 
significant risks associated with the misuse of this 
approach.

%% file: tables/ours_distributed_context_complexity.tex
\begin{table*}[hbt]
    \centering
    \scriptsize
    \setlength{\tabcolsep}{2.8pt}
    \begin{tabular}{
        l
        >{\columncolor{gray!20}}r >{\columncolor{gray!20}}r >{\columncolor{gray!20}}r >{\columncolor{gray!20}}r
        c
        >{\columncolor{white}}r >{\columncolor{white}}r >{\columncolor{white}}r >{\columncolor{white}}r
        c
        >{\columncolor{gray!20}}r >{\columncolor{gray!20}}r >{\columncolor{gray!20}}r >{\columncolor{gray!20}}r
        c
        >{\columncolor{white}}r >{\columncolor{white}}r >{\columncolor{white}}r >{\columncolor{white}}r
    }

        \toprule
        \multirow{3}{*}{\textbf{Dataset}} & \multicolumn{4}{>{\columncolor{gray!20}}c}{\textbf{Repository-level}} && \multicolumn{4}{c}{\textbf{File-level}} && \multicolumn{4}{>{\columncolor{gray!20}}c|}{\textbf{Self-contained}}  && \multicolumn{4}{c}{\textbf{Total}}  \\
        \cmidrule{2-5} \cmidrule{7-10} \cmidrule{12-15} \cmidrule{17-20}
        & \#NL & \#GT & Cyclo. & \#Funcs. && \#NL & \#GT & Cyclo. & \#Funcs. && \#NL & \#GT & Cyclo. & \#Funcs. &&
        \#NL & \#GT & Cyclo. & \#Funcs.\\
        \midrule
astropy & 360.6 &  447.9 & 10.5 & 51 &&  404.1 &   195.7 & 4.9 & 14 &&  227.6 &  238.2  & 5.9 & 20  &&  336.5 &  357.0  & 8.5  & 85\\
datasets & 499.7 &  348.7 & 9.6 & 19 &&  292.3 &   168.8 & 5.1 & 18 &&  312.5 &  128.9  & 3.1 & 22  &&  366.6 &  211.9  & 5.8  & 59\\
flask & 277.9 &  171.9 & 5.2 & 8 &&  356.1 &   160.4 & 5.2 & 10 &&  233.0 &  88.6  & 3.5 & 25  &&  270.0 &  120.8  & 4.2  & 43\\
more-itertools & 224.6 &  70.2 & 2.6 & 8 &&  293.4 &   127.0 & 4.0 & 18 &&  243.2 &  105.2  & 5.1 & 60  &&  252.0 &  106.5  & 4.6  & 86\\
plotly.py & 1,357.3 &  1373.9 & 37.0 & 15 &&  1,337.3 &   356.0 & 7.9 & 7 &&  1,687.3 &  590.5  & 24.3 & 54  &&  1,589.9 &  723.5  & 25.3  & 76\\
pylint & 180.9 &  386.2 & 14.0 & 9 &&  163.9 &   272.3 & 7.1 & 7 &&  169.8 &  178.9  & 8.0 & 10  &&  172.0 &  275.8  & 9.8  & 26\\
scikit-learn & 217.9 &  372.1 & 7.2 & 237 &&  299.7 &   294.9 & 5.6 & 27 &&  212.8 &  237.1  & 5.5 & 50  &&  224.1 &  344.0  & 6.8  & 314\\
seaborn & 230.8 &  538.8 & 15.0 & 13 &&  272.7 &   168.0 & 4.8 & 38 &&  163.3 &  201.1  & 5.7 & 27  &&  227.9 &  241.3  & 6.8  & 78\\
sphinx & 201.1 &  548.4 & 15.4 & 14 &&  228.0 &   240.8 & 5.5 & 4 &&  274.1 &  88.1  & 3.9 & 15  &&  237.5 &  301.9  & 8.9  & 33\\
sympy & 874.7 &  603.4 & 17.8 & 67 &&  821.3 &   182.7 & 6.5 & 14 &&  962.4 &  138.7  & 4.6 & 16  &&  881.5 &  466.0  & 14.0  & 97\\
xarray & 791.5 &  366.9 & 10.7 & 57 &&  731.6 &   102.1 & 2.8 & 20 &&  306.2 &  114.3  & 2.3 & 6  &&  742.0 &  284.8  & 8.2  & 83\\
\midrule
Average &  431.9 &  441.2  & 10.8 & 498 &&  428.1 &  192.0  & 5.0 & 177 &&  528.0 &  233.8  & 8.3 & 305 &&  461.1 &  331.6  & 9.0 & 980 \\
        \bottomrule
    \end{tabular}
    \caption{Basic statistics of \ours, with details broken down by each collected repository.}
    \label{tab:ours_complexity}
\end{table*}

%% file: tables/eval_result_per_repo.tex
\begin{table*}[htb]
    \centering
    \scriptsize
    \setlength{\tabcolsep}{3.6pt}
    \begin{tabular}{l rrrrrrrrrrr | r}
        \toprule
\textbf{Setting} & \textbf{pylint} & \textbf{sympy} & \textbf{sphinx} & \textbf{seaborn} & \textbf{flask} & \textbf{more-itertools} & \textbf{scikit-learn} & \textbf{xarray} & \textbf{datasets} & \textbf{plotly.py} & \textbf{astropy} & \textbf{Total} \\
    \midrule
    CodeLlama-7B & 7.69 &7.22 &9.09 &23.08 &37.21 &13.95 &5.10 &6.02 &22.03 &6.58 &5.88 &10.41 \\
    CodeLlama-34B & 0.00 &4.12 &18.18 &21.79 &32.56 &24.42 &5.73 &16.87 &18.64 &14.47 &10.59 &12.76 \\
    DeepSeekCoder-6.7B & 7.69 &6.19 &15.15 &26.92 &41.86 &12.79 &6.37 &16.87 &16.95 &25.00 &14.12 &14.08 \\
    DeepSeekCoder-33B & 3.85 &7.22 &27.27 &28.21 &37.21 &11.63 &7.32 &15.66 &33.90 &43.42 &16.47 &17.14 \\
    OpenCodeInterpreter-6.7B & 3.85 &4.12 &12.12 &16.67 &41.86 &18.60 &6.37 &10.84 &16.95 &21.05 &12.94 &12.45 \\
    OpenCodeInterpreter-33B & 11.54 &7.22 &21.21 &20.51 &41.86 &29.07 &8.28 &13.25 &25.42 &23.68 &16.47 &16.33 \\
    \midrule
    Claude 3.5 Sonnet & 19.23 &11.34 &15.15 &25.64 &39.53 &38.37 &6.05 &9.64 &27.12 &36.84 &10.59 &17.45 \\
    DeepSeek-V2.5 & 15.38 &13.40 &27.27 &30.77 &41.86 &36.05 &9.87 &16.87 &30.51 &35.53 &16.47 &20.71 \\
    GPT-4o-Mini & 11.54 &11.34 &15.15 &25.64 &39.53 &18.60 &6.05 &10.84 &33.90 &25.00 &9.41 &15.00 \\
    GPT-4o & 19.23 &15.46 &27.27 &37.18 &58.14 &43.02 &13.69 &24.10 &47.46 &44.74 &23.53 &27.04 \\
    \bottomrule
    \end{tabular}
    \caption{Pass@1(\%) of LLMs on \ours under RAG\textsubscript{Dense} retrieval, with details shown for each repository. }
    \label{tab:eval_result_by_repo}
\end{table*}

%% file: tables/retrieval_stat_by_repo.tex
\begin{table}[ht]
    \centering
    \scriptsize
    \setlength{\tabcolsep}{20pt}
    \begin{tabular}{l rr}
    \toprule
    \textbf{Repository} & \textbf{RAG\textsubscript{BM25}} & \textbf{RAG\textsubscript{Dense}} \\
    \midrule
    pylint & 0.14 & 0.18 \\
    sympy & 0.16 & 0.14 \\
    sphinx & 0.09 & 0.11 \\
    seaborn & 0.17 & 0.21 \\
    flask & 0.11 & 0.20 \\
    more-itertools & 0.05 & 0.10 \\
    scikit-learn & 0.09 & 0.07 \\
    xarray & 0.07 & 0.15 \\
    datasets & 0.21 & 0.24 \\
    plotly.py & 0.02 & 0.11 \\
    astropy & 0.17 & 0.15 \\
    \midrule
    \textbf{Total} & 0.12 & 0.15 \\
    \bottomrule
    \end{tabular}
    \caption{Recall rate of the top three and top ten retrieved contents across retrieval settings for repository-level and file-level contexts.}
    \label{tab:retrieval-stat-by-repo}
\end{table}

%% file: tables/inference_len_vs_correct_rate_dense.tex
        
\begin{table}[t]
    \centering
    \scriptsize
    \setlength{\tabcolsep}{5pt}
    \begin{tabular}{l >{\columncolor{gray!20}}rr c >{\columncolor{gray!20}}rr c >{\columncolor{gray!20}}rr}
        \toprule
        \multirow{3}{*}{\textbf{Models}} & \multicolumn{2}{c}{\textbf{RAG\textsubscript{BM25}}} && \multicolumn{2}{c}{\textbf{RAG\textsubscript{Dense}}}  && \multicolumn{2}{c}{\textbf{Current-File}} \\
        \cmidrule{2-3} \cmidrule{5-6} \cmidrule{8-9}
        & Pass & Fail &&Pass & Fail &&Pass & Fail \\
        \midrule
CodeLlama-7B &  18.2  & 11.6  && 9.1  & 11.9  && 30.4  & 22.3  \\
CodeLlama-34B &  1.8  & 15.2  && 1.9  & 11.5  && 22.5  & 34.5  \\
DeepSeek-6.7B &  2.2  & 9.6  && 7.0  & 5.7  && 20.0  & 13.1  \\
DeepSeek-33B &  8.2  & 7.4  && 8.4  & 3.9  && 17.9  & 26.5  \\
OCI-6.7B &  2.3  & 2.5  && 3.5  & 2.8  && 2.5  & 1.8  \\
OCI-33B &  2.4  & 1.7  && 2.1  & 1.5  && 2.3  & 1.3  \\
\midrule
Claude-3.5 &  19.2  & 1.7  && 25.5  & 1.7  && 9.8  & 7.0  \\
DeepSeek-v2.5 &  7.3  & 6.4  && 8.0  & 7.6  && 3.6  & 1.7  \\
GPT-4-Mini &  4.0  & 1.9  && 4.9  & 1.8  && 12.1  & 1.5  \\
GPT-4 &  1.7  & 1.5  && 1.7  & 1.5  && 2.5  & 1.2  \\
        \bottomrule
    \end{tabular}
    \caption{Ratio of generation length over canonical solution length}
    \label{tab:inference-len-dense-unfiltered}
\end{table}

\begin{table}[t]
    \centering
    \scriptsize
    \setlength{\tabcolsep}{5pt}
    \begin{tabular}{l >{\columncolor{gray!20}}rr c >{\columncolor{gray!20}}rr c >{\columncolor{gray!20}}rr}
        \toprule
        \multirow{3}{*}{\textbf{Models}} & \multicolumn{2}{c}{\textbf{RAG\textsubscript{BM25}}} && \multicolumn{2}{c}{\textbf{RAG\textsubscript{Dense}}}  && \multicolumn{2}{c}{\textbf{Current-File}} \\
        \cmidrule{2-3} \cmidrule{5-6} \cmidrule{8-9}
        & Pass & Fail &&Pass & Fail &&Pass & Fail \\
        \midrule
CodeLlama-7B &  0.9  & 1.7  && 1.0  & 1.7  && 1.0  & 1.8  \\
CodeLlama-34B &  0.9  & 1.5  && 0.8  & 1.4  && 0.9  & 1.5  \\
DeepSeek-6.7B &  1.0  & 1.6  && 0.9  & 1.4  && 0.9  & 1.8  \\
DeepSeek-33B &  0.8  & 1.8  && 1.0  & 1.7  && 0.9  & 1.9  \\
OCI-6.7B &  1.6  & 2.1  && 2.8  & 2.4  && 1.7  & 1.5  \\
OCI-33B &  1.7  & 1.2  && 1.4  & 1.1  && 1.4  & 1.0  \\
\midrule
Claude-3.5 &  1.4  & 1.4  && 1.6  & 1.4  && 1.4  & 2.1  \\
DeepSeek-v2.5 &  1.2  & 1.2  && 1.1  & 1.3  && 1.0  & 1.4  \\
GPT-4-Mini &  1.4  & 1.1  && 1.5  & 1.0  && 9.1  & 1.0  \\
GPT-4 &  1.3  & 1.3  && 1.3  & 1.1  && 0.9  & 1.0  \\
        \bottomrule
    \end{tabular}
    \caption{Ratio of filtered generation length over canonical solution length.}
    \label{tab:inference-len-filtered-dense}
\end{table}

%% file: tables/pass_vs_callee_len.tex
\begin{table}[t]
    \centering
    \scriptsize
    \setlength{\tabcolsep}{4pt}
    \begin{tabular}{l l  >{\columncolor{gray!20}}r r  >{\columncolor{gray!20}}r r  >{\columncolor{gray!20}}r}
        \toprule
        
    \multirow{3}{*}{\textbf{Model}} & \multirow{3}{*}{\textbf{Retrieval}} & \multicolumn{5}{c}{\textbf{\# of Dependencies}} \\
    \cmidrule{3-7}
    & & 0 & 1 & 2 & [3,4] & [5,35] \\
        \midrule

\multirow{3}{*}{DeepSeek-v2.5}  
        & RAG\textsubscript{BM25} & 31.8 & 22.1 & 14.2 & 7.1 & 5.7 \\
        & RAG\textsubscript{Dense} & 35.4 & 28.4 & 14.2 & 6.0 & 6.2 \\
        & Current-File & 42.6 & 31.1 & 22.0 & \textbf{17.3} & 9.1 \\
        \midrule
\multirow{3}{*}{Claude-3.5}  
        & RAG\textsubscript{BM25} & 27.2 & 18.9 & 9.9 & 2.4 & 2.3 \\
        & RAG\textsubscript{Dense} & 33.1 & 22.6 & 12.8 & 4.8 & 0.6 \\
        & Current-File & 25.2 & 28.4 & 16.3 & 14.9 & 8.5 \\
        \midrule
\multirow{3}{*}{GPT-4-Mini}  
        & RAG\textsubscript{BM25} & 22.6 & 20.5 & 14.9 & 6.0 & 5.1 \\
        & RAG\textsubscript{Dense} & 24.3 & 21.1 & 13.5 & 5.4 & 2.8 \\
        & Current-File & 30.5 & 23.2 & 14.9 & 8.9 & 5.7 \\
        \midrule
\multirow{3}{*}{GPT-4}  
        & RAG\textsubscript{BM25} & 39.3 & 32.6 & 22.7 & 16.7 & \textbf{14.8} \\
        & RAG\textsubscript{Dense} & \textbf{44.6} & \textbf{36.3} & 16.3 & 12.5 & 9.1 \\
        & Current-File & 39.3 & 33.7 & \textbf{24.1} & 16.1 & 10.2 \\
        \bottomrule
    \end{tabular}
    \caption{Pass Rate by Number of dependencies, binned by the number of unique dependencies. The bins are distributed as follows: 0 (305 instances), 1 (190 instances), 2 (141 instances), [3,4] (168 instances), and [5,35] (176 instances).}
    \label{tab:pass-rate-vs-callees-length}
\end{table}

%% file: tables/lite_result.tex
\begin{table}[t]
    \centering
    \scriptsize
    \setlength{\tabcolsep}{6pt}
    \begin{tabular}{lrrr}
        \toprule
        {\textbf{Models}} & {\textbf{RAG\textsubscript{BM25}}} &{\textbf{RAG\textsubscript{Dense}}} & {\textbf{Current-File}}  \\
        \midrule
        CodeLlama-7B & 1.0 & 1.5 & 1.0 \\
        CodeLlama-34B & 2.5 & 2.0 & 1.0 \\
        DeepSeek-6.7B & 2.5 & 2.0 & 0.5 \\
        DeepSeek-33B & 4.0 & 3.5 & 2.5 \\
        OpenCodeInterpreter-6.7B & 1.0 & 5.0 & 1.5 \\
        OpenCodeInterpreter-33B & 4.0 & 6.0 & 5.5 \\
        \midrule
        Claude-3.5 & 2.5 & 3.0 & 2.5 \\
        DeepSeek-v2.5 & 4.0 & 7.5 & 5.5 \\
        GPT-4o-Mini & 8.0 & 6.5 & 2.5 \\
        GPT-4o & 9.0 & 8.5 & 4.5 \\
        \bottomrule
    \end{tabular}
    \caption{Pass@1(\%) of SOTA LLMs on \ours{-LITE}.}
    \label{tab:lite-pass-rates}
\end{table}

%% file: tables/ous_lite_distributed_context_complexity.tex
\begin{table*}[hbt]
    \centering
    \scriptsize
    \setlength{\tabcolsep}{2pt}
    \begin{tabular}{
        l
        >{\columncolor{gray!20}}r >{\columncolor{gray!20}}r >{\columncolor{gray!20}}r >{\columncolor{gray!20}}r
        c
        >{\columncolor{white}}r >{\columncolor{white}}r >{\columncolor{white}}r >{\columncolor{white}}r
        c
        >{\columncolor{gray!20}}r >{\columncolor{gray!20}}r >{\columncolor{gray!20}}r >{\columncolor{gray!20}}r
        c
        >{\columncolor{white}}r >{\columncolor{white}}r >{\columncolor{white}}r >{\columncolor{white}}r
    }

        \toprule
        \multirow{3}{*}{\textbf{Dataset}} & \multicolumn{4}{>{\columncolor{gray!20}}c}{\textbf{Repository-level}} && \multicolumn{4}{c}{\textbf{File-level}} && \multicolumn{4}{>{\columncolor{gray!20}}c|}{\textbf{Self-contained}}  && \multicolumn{4}{c}{\textbf{Total}}  \\
        \cmidrule{2-5} \cmidrule{7-10} \cmidrule{12-15} \cmidrule{17-20}
        & \#NL & \#GT & Cyclo. & \#Funcs. && \#NL & \#GT & Cyclo. & \#Funcs. && \#NL & \#GT & Cyclo. & \#Funcs. &&
        \#NL & \#GT & Cyclo. & \#Funcs.\\
        \midrule
astropy & 321.7 &  1,104.0 & 25.3 & 3 &&  678.8 &   1,107.2 & 20.0 & 4 &&  485.3 &  618.8  & 14.0 & 9  &&  503.0 &  831.9  & 17.6  & 16\\
datasets & 1,432.8 &  832.8 & 24.0 & 4 &&  1,331.0 &   526.0 & 14.0 & 1 &&  1211.0 &  318.0  & 8.0 & 1  &&  1,378.8 &  695.8  & 19.7  & 6\\
flask & 661.0 &  424.0 & 12.0 & 1 &&  595.0 &   515.0 & 12.0 & 1 &&  452.0 &  691.0  & 16.0 & 1  &&  569.3 &  543.3  & 13.3  & 3\\
more-itertools & - &  - & - & 0 &&  503.0 &   815.0 & 11.0 & 1 &&  - &  -  & - & 0  &&  503.0 &  815.0  & 11.0  & 1\\
plotly.py & 1,806.0 &  3,393.0 & 132.0 & 1 &&  1,574.4 &   1,191.8 & 30.2 & 14 &&  1,933.0 &  1464.2  & 56.9 & 23  &&  1,797.5 &  1,414.6  & 49.0  & 38\\
pylint & - &  - & - & 0 &&  - &   - & - & 0 &&  385.0 &  762.0  & 16.0 & 1  &&  385.0 &  762.0  & 16.0  & 1\\
scikit-learn & 420.3 &  886.3 & 15.5 & 51 &&  559.8 &   727.4 & 8.0 & 5 &&  493.6 &  713.0  & 15.2 & 5  &&  437.8 &  859.1  & 14.8  & 61\\
seaborn & 404.3 &  750.0 & 18.3 & 3 &&  - &   - & - & 0 &&  363.8 &  678.2  & 22.5 & 4  &&  381.1 &  709.0  & 20.7  & 7\\
sphinx & - &  - & - & 0 &&  - &   - & - & 0 &&  418.0 &  946.0  & 17.0 & 1  &&  418.0 &  946.0  & 17.0  & 1\\
sympy & 978.0 &  424.8 & 10.8 & 4 &&  1,135.2 &   1,155.1 & 34.5 & 24 &&  810.2 &  496.1  & 14.4 & 16  &&  1,002.7 &  849.1  & 25.0  & 44\\
xarray & - &  - & - & 0 &&  1,562.5 &   695.1 & 18.5 & 17 &&  1,158.8 &  367.6  & 13.0 & 5  &&  1,470.7 &  620.6  & 17.2  & 22\\
\midrule
\textbf{Total} &  533.2 &  889.7  & 17.9 & 67 &&  1,250.6 &  987.3  & 25.7 & 67 &&  1,120.8 &  879.0  & 29.6 & 66 &&  967.4 &  918.9  & 24.39 & 200 \\
        \bottomrule
    \end{tabular}
    \caption{Basic statistics of \ours{\_LITE}, with details broken down by each collected repository.}
    \label{tab:ours_complexity_by_repo_lite}
\end{table*}